\documentclass[11pt,a4paper]{article}
\pdfoutput=1
\usepackage{jheppub}
\usepackage{amsmath}
\usepackage{epsfig}
\usepackage{amssymb}
\usepackage{graphics}
\usepackage[active]{srcltx}
\usepackage{epstopdf}
\usepackage{pdfsync}
\usepackage{shuffle}
\usepackage{slashed}
\usepackage{soul}

\usepackage{tikz}
\usetikzlibrary{decorations.pathmorphing}
\usetikzlibrary{arrows.meta}

\def\spaa #1{\langle #1\rangle}

\def\I{\tiny \mbox{I}}
\def\II{\tiny \mbox{II}}

\title{Four-loop Anomalous Dimensions of Scalar-QED Theory from Operator Product Expansion}
\author[a]{Rijun Huang}
\author[b]{, Qingjun Jin}
\author[c]{and Yi Li}

\affiliation[a]{School of Physics and Technology, Nanjing Normal University, No.1 Wenyuan Road, Nanjing 210046, P.R.China}
\affiliation[b]{Graduate School of China Academy of Engineering Physics, No. 10 Xibeiwang East Road, Haidian District, Beijing, 100193, P.R.China}
\affiliation[c]{School of Physics, Hubei University, No.368 Youyi Road, Wuhan 430062, P.R.China}

\emailAdd{huang@njnu.edu.cn}
\emailAdd{qjin@gscaep.ac.cn}
\emailAdd{yili@gscaep.ac.cn}

\abstract{We apply the Operator Product Expansion (OPE) algorithm to the renormalization of scalar-QED theory, with a specific focus on the fixed-charge operator $\phi^Q$. Within the OPE framework, the anomalous dimension of the $\phi^Q$ operator is perturbatively computed to four-loop order in the modified minimal subtraction scheme, extending beyond the previously available three-loop result. The beta functions, as well as the mass and field anomalous dimensions, are also computed at this order. An alternative loop-integrand construction method is proposed, based on graph decomposition and skeleton expansion techniques, for deriving the integrands of one-Particle-Irreducible correlation functions. This work represents the first non-trivial validation of the OPE algorithm for higher-loop renormalization beyond pure scalar theories. The present successful computations further confirm the efficiency and versatility of the OPE algorithm in renormalization analysis.}


\begin{document}

\maketitle \flushbottom

%%%%%%%%%%%%%%%%
\section{Introduction}
\label{sec:introduction}
%%%%%%%%%%%%%%%%%%%

Scalar Quantum Electrodynamics (scalar-QED) stands as one of the most foundational gauge theories and occupies an important position in the development of quantum field theory. Its study is closely related to the evolution of quantum electrodynamics
%, which was originally formulated to describe electron-photon interactions and generalize electromagnetic phenomena within the framework of quantum mechanics 
\cite{dirac1927quantum}. By introducing scalar particles instead of spin-half fermions into this theoretical framework, scalar-QED enables the systematic investigation of interactions between scalar fields and photons, thus establishing itself as an ideal theoretical platform for interpreting a broad range of physical phenomena
across diverse energy scales and research domains \cite{Zinn:2021QFT,Shankar:2017zag}. For example, scalar-QED is also known as the Abelian Higgs model, which is intimately associated with the Higgs mechanism \cite{Anderson:1963pc,Englert:1964et,Higgs:1964pj,Guralnik:1964eu} in elementary particle physics. %, in which massive scalar and vector particles are generated via spontaneous symmetry breaking \cite{PhysRevD.7.1888}. 
Furthermore, the non-relativistic limit of scalar-QED gives rise to the well-known Ginzburg-Landau model \cite{Ginzburg:1950sr}, which is phenomenologically relevant to a wide range of experimental effects in condensed matter physics \cite{Hohenberg:2015jgf}. These effects include phase transition phenomena in superconductors and liquid crystals \cite{Halperin:1973jh,Herbut:1996ut}, quantum Hall effects \cite{PhysRevLett.70.1501}, non-equilibrium pattern formation \cite{RevModPhys.65.851}, and more. %Beyond these areas, the potential applications of scalar-QED theory to broader research scenarios have been extensively explored. 
Researches beyond these areas have also been extensively explored, for instance in the study of cosmic strings in the early universe \cite{Kibble:1976sj,Hindmarsh:2008dw,Dufaux:2010cf,Hindmarsh:2017qff,Hindmarsh:2021mnl}, and in the worldline instanton technique for semiclassical non-perturbative quantum field theory calculations \cite{Affleck:1981bma,Dunne:2005sx}.

Many of the aforementioned applications are related to the phase transitions and critical phenomena in diverse physical systems. Such research relies on the renormalization group (RG) computations to determine critical points and critical exponents \cite{Wilson:1971dc,PELISSETTO2002549,Moshe:2003xn}, which can then be compared with experimental data. However, perturbative RG calculations for scalar-QED theory (or the Abelian Higgs model) remain quite limited. The one-loop and two-loop calculations were conducted long ago \cite{Halperin:1973jh,Vladimirov:1979ib,vanDamme:1982bg,PhysRevB.41.4083,Herbut:1996ut}, whereas a four-loop renormalization computation was only recently completed 
%for scalar-QED theory describing $n$ complex scalars coupled to a fluctuating $U(1)$ gauge field 
\cite{Ihrig:2019kfv}, in which the beta functions and anomalous dimensions of the fields were derived. In contrast, as of the time of writing, the state-of-the-art results for scalar field theories are two to four loops higher than those of scalar-QED \cite{Batkovich:2016jus,Kompaniets:2016hct,Kompaniets:2017yct,Schnetz:2016fhy,Schnetz:2025opm,Gracey:2015tta,Kompaniets:2021hwg,Borinsky:2021jdb,Schnetz:2025wtu}.
The discrepancy between renormalization of scalar-QED and scalar field theories becomes even more pronounced regarding the composite operators. A specific class of composite operators is the fixed-charge symmetric traceless operator $\phi^Q$, whose anomalous dimensions for certain values of $Q$ are relevant to second-order phase transitions \cite{DePrato:2003yd,Calabrese:2004ca} observed in various statistical physics systems. In recent years, progress in the renormalization computation of the $\phi^Q$ operator has been driven by both semiclassical approaches \cite{Arias-Tamargo:2019xld,Badel:2019oxl,Antipin:2020abu,Giombi:2020enj,Arias-Tamargo:2020fow,Antipin:2021jiw,Giombi:2022gjj, Antipin:2022naw,Antipin:2022hfe,Antipin:2023tar} and perturbative methods \cite{Jack:2021ypd,Jin:2022nqq,Bednyakov:2022guj,Jack:2021ziq,Huang:2024hsn,Huang:2025rdy}. The semiclassical method enables the renormalization of the $\phi^Q$ operator to arbitrary loop order at the large charge limit, but the current calculations for scalar-QED theory are limited to the leading and next-to-leading orders in the large charge expansion series \cite{Antipin:2022hfe}, leaving a gap in obtaining complete higher loop order predictions. On the perturbative side, the complete multi-loop results could serve as an independent cross-check with semiclassical outcomes, and facilitate general discussions on charge dependence. But for scalar-QED theory the most recent perturbative result of the $\phi^Q$ operator is the three-loop one derived in the same paper \cite{Antipin:2022hfe}, serving as a comparison with the semiclassical findings. In contrast, for scalar theories the renormalization of the $\phi^Q$ operator has been perturbatively computed up to six loops for quartic interactions \cite{Bednyakov:2022guj}, and five loops for cubic interactions \cite{Huang:2025rdy}. In this paper, we advance the perturbative front by extending the four-loop renormalization of scalar-QED theory to the $\phi^Q$ operator.

Perturbative computation of renormalization functions primarily poses two key challenges. One is the evaluation of Ultra-Violent (UV) divergences in individual Feynman integrals, and the other is the non-trivial summation of UV divergences from Feynman graphs that contribute to these renormalization functions. For the former, the graphical function method \cite{Schnetz:2013hqa,Golz:2015rea,Borinsky:2021gkd,Schnetz:2024qqt,HP} is a powerful tool which enables the systematic evaluation of scalar Feynman integrals even up to 8-loops \cite{Schnetz:2016fhy,Schnetz:2025opm,Schnetz:2025wtu,Huang:2025ree}. However for scalar-QED we need to consider Feynman integrals with complicated numerator structures. Conventionally the Integration-By-Parts (IBP) method \cite{Tkachov:1981wb,Chetyrkin:1981qh,Laporta:2000dsw} is employed to deal with these integrals. The efficiency of integral reduction algorithm ({\sl e.g.}, \cite{Smirnov:2008iw,Maierhofer:2017gsa,Wu:2023upw,Guan:2024byi}) and the availability of pre-evaluated master integrals constrain the loop orders that can achieve. For the latter, a suite of powerful approaches has been developed to manage the UV divergences. Methodologies such as Infra-Red (IR) rearrangement \cite{Vladimirov:1979zm,Chetyrkin:1980pr,Caswell:1981ek}, the framework of the $R$-operation (and its more advanced variant, the $R^\ast$-operation) \cite{Chetyrkin:1982nn,CHETYRKIN1984419,Larin:2002sc}, and the massive vacuum bubble method \cite{Misiak:1994zw,Chetyrkin:1997fm} have been instrumental in organizing the UV divergences arising from large ensembles of Feynman graphs. However, the aforementioned methods are mostly well-developed and efficient for correlation functions with a small number of external points (typically no more than four. But certainly they can be applied to compute correlation functions beyond four-points with some sophisticated techniques, for instance in \cite{Cao:2021cdt,Henriksson:2025hwi,Henriksson:2025vyi}). When considering the perturbative renormalization of composite operators, for instance the $\phi^Q$ operator, multi-leg correlation functions involving operator insertion and $Q$ external fields should be considered. As $Q$ can be very large, it makes the $\phi^Q$ problem exceptionally challenging. To address this limitation and generalize high-loop renormalization to generic composite operators, a promising method based on the Operator Product Expansion (OPE) has been systematically developed \cite{Huang:2024hsn}. The OPE has long been recognized as deeply intertwined with the anomalous dimensions of operators (see {\sl e.g.} \cite{Collins_1984}), inspiring the idea of extracting these fundamental quantities from the Wilson coefficients of the expansion \cite{Collins_1984,Eden:2012fe,Prochazka:2019fah}. This concept has been refined into a systematic algorithm for the renormalization of generic composite operators in \cite{Huang:2024hsn}, addressing the above-mentioned two challenges at the same time. With the proposed OPE algorithm, the renormalization of the $\phi^Q$ operator for six-dimensional cubic scalar theory has been pushed to five loop order \cite{Huang:2025rdy}, confirming its efficiency and versatility.

In this study, we extend the application of the OPE-based algorithm to the four-loop renormalization of the $\phi^Q$ operator in four-dimensional scalar-QED theory. This paper is structured as follows. In \S\ref{sec:review}, we present the definition of the scalar-QED theory and the renormalization factors considered in this work, and briefly review the OPE algorithm for renormalization. In \S\ref{sec:integrand}, we describe the primitive diagram method for constructing loop integrands. This method is built upon graph decomposition and skeleton expansion techniques, and has been optimized to meet the demands of high-loop integrand construction in this study. In \S\ref{sec:result}, we rephrase the proposed OPE algorithm in the context of scalar-QED theory and use it to computed the four-loop anomalous dimensions of the $\phi^Q$ operator. The four-loop result and a discussion on the gauge dependence of the anomalous dimensions are presented. Conclusion and discussion are provided in \S\ref{sec:conclusion}, and a subtle point regarding the computation of four-scalar correlation functions is detailed in Appendix \ref{appendix:1PR}.

%%%%%%%%%%%%%%%%
\section{The theory and the methods}
\label{sec:review}
%%%%%%%%%%%%%%%%%%%

%%%%%%%%%%%%%%%%
\subsection{The scalar-QED theory}
\label{subsec:review-sQED}
%%%%%%%%%%%%%%%%%%%

We investigate the renormalization of scalar-QED theory describing a complex scalar coupled to a gauge field. In Euclidean space, its Lagrangian is given by 
\begin{equation}
\mathcal{L}_{\tiny\mbox{sQED}}=\frac{1}{4}F_{\mu\nu}F^{\mu\nu}+(D_\mu \phi)^{\dagger}D^\mu \phi +m^2\phi\bar{\phi}+\frac{\lambda}{4}(\bar{\phi}\phi)^2~,
\end{equation}
where $\phi$ denotes the complex scalar field, $F_{\mu\nu}=\partial_\mu A_\nu-\partial_\nu A_\mu$ with $A_\mu$ being the $U(1)$ gauge field, and the covariant derivative is defined as $D_\mu:=\partial_\mu-ieA_\mu$. We adopt the standard textbook Feynman rules for propagators and vertices derived from this Lagrangian, and the $R_{\xi}$-gauge is employed for the photon propagator. When necessary, one can simply set $\xi=0$ for the Landau gauge and $\xi=1$ for the Feynman gauge. All perturbative computations are performed within the modified minimal subtraction ($\overline{\mbox{MS}}$) regularization scheme \cite{tHooft:1973mfk} in $d=(4-2\epsilon)$-dimensions.

Scalar-QED is renormalizable in four dimensions. Following standard textbook definitions, we introduce the renormalized vector and scalar fields as
\begin{equation}
A_{\mu,0}=Z_{A}^{\frac{1}{2}}A_\mu~~~,~~~\phi_0=Z_{\phi}^{\frac{1}{2}}\phi~,
\end{equation}
and the renormalized coupling constants as
\begin{equation}
e_0=\mu^{\epsilon}\widetilde{Z}_{e}e~~~,~~~\lambda_0=\mu^{2\epsilon}\widetilde{Z}_\lambda\lambda~~~
\mbox{with}~~~\widetilde{Z}_{e}=Z_{A}^{-\frac{1}{2}}~~~,~~~\widetilde{Z}_{\lambda}=Z_{\lambda}Z_{\phi}^{-2}~,
\end{equation}
where $A_{\mu,0},\phi_0$ and $e_0,\lambda_0$ represent the bare fields and bare coupling factors, respectively, and $\mu$ is the scale parameter. From gauge invariance, we have $Z_AZ_e=1$. For the renormalization analysis, we need to compute the field renormalization factor $Z_\phi$ and $Z_A$, as well as the coupling constant renormalization factor $Z_\lambda$. The beta functions and field anomalous dimensions can be calculated from these $Z$-factors via standard textbook definitions. Similarly, the mass anomalous dimension can be derived from the $Z_m$-factor of the gauge-invariant operator $\widehat{O}_m:=\bar{\phi}\phi$. The primary focus of this paper is the fixed-charge operator $\phi^Q$, whose renormalized form is defined as $\phi^{Q}_{R}=Z_{\phi^Q}\phi_0^Q$. Its anomalous dimension $\gamma_{\phi^Q}$ and scaling dimension $\Delta_Q$ are defined as
\begin{equation}
\gamma_{\phi^Q}:=-\frac{\partial{\ln Z_{\phi^Q}}}{\partial \ln \mu}~~~,~~~\Delta_Q=\left(\frac{d}{2}-1\right)Q+\gamma_{\phi^Q}~.
\end{equation}
All these $Z$-factors can be computed using the OPE algorithm for renormalization \cite{Huang:2024hsn}, which is reviewed in the next subsection.

%%%%%%%%%%%%%%%%
\subsection{OPE algorithm for renormalization}
\label{subsec:review-OPE}
%%%%%%%%%%%%%%%%%%%

In the perturbative renormalization of generic composite operators, the evaluation of the required correlation functions presents a substantial difficulty, particularly for those with a large number of external legs. The OPE-based algorithm proposed in \cite{Huang:2024hsn} is designed for such situations. This algorithm constitutes a non-trivial generalization of the deep connection between the Wilson coefficients of the OPE and the anomalous dimensions of operators. Within the OPE framework, the UV divergences required for renormalization are derived from two-point propagator-type integrals, which are deduced from the original multi-leg correlation functions by removing external legs. This transformation significantly simplifies the calculation of UV divergences, and is particularly advantageous for composite operators involving correlation functions with numerous external legs. Furthermore, UV divergences are treated globally during evaluation, and the trivial summation over all two-point integrals yields the UV divergences necessary for renormalization, eliminating the need for cumbersome subtraction of sub-divergences. These advantages facilitate multi-loop renormalization computations in two respects. Firstly, they simplify the evaluation of loop integrals for UV divergences and enable the renormalization of generic composite operators with numerous external legs. Secondly, the summation of UV divergences can be performed systematically, which allows for the automatic implementation of the OPE algorithm in modern computers. Importantly, the generality of the OPE ensures the applicability of the algorithm to diverse scenarios, including scalar theories, theories involving particles with spins, and operators with (higher) derivative or Lorentz indices, among others. This opens up possibilities for computing the anomalous dimensions of complicated operators and has the potential to push the precision of perturbative renormalization to the next loop order. The OPE also demonstrates advantages for renormalization in condensate calculations \cite{Marino:2024uco,Liu:2025bqq}.

The derivation of the OPE algorithm begins with the general OPE of two operators $\widehat{O}_{\I}$, $\widehat{O}_{\II}$ in momentum space, given by
\begin{equation}
\widehat{O}_{\II}^{\tiny\mbox{ren}}(k-p)\widehat{O}_{\I}^{\tiny\mbox{ren}}(p)~\xrightarrow{p\to\infty}~\sum_{\mathcal{O}} C_{\mathcal{O}}^{\tiny\mbox{ren}}(p)\widehat{\mathcal{O}}^{\tiny\mbox{ren}}(k)~,\label{eqn:OPE-ren}
\end{equation}
where the superscript $^{\tiny\mbox{ren}}$ emphasizes renormalized operators in dimensional regularization. The expansion basis $\widehat{\mathcal{O}}$'s form a set of composite operators ordered by their dimensions. The Wilson coefficients $C_{\mathcal{O}}^{\tiny\mbox{ren}}(p)$ are functions of all coupling constants and the large momentum parameter $p^2$, which are generally singular in $p^2$. The expansion is performed in the $p\to\infty$ limit, and the large momentum expansion technique can be employed to analyze the asymptotic behaviors \cite{Smirnov:2002pj}\footnote{At this limit, the two operators can be considered as two hard fields propagating in the background of soft fields, as described in \cite{Huang:2016bmv,Huang:2022qnx}.}. Alternatively, applying the OPE to two bare operators yields 
\begin{equation}
\widehat{O}_{\II}^{\tiny\mbox{bare}}(k-p)\widehat{O}_{\I}^{\tiny\mbox{bare}}(p)~\xrightarrow{p\to\infty}~ \sum_{\mathcal{O}} C_{\mathcal{O}}^{\tiny\mbox{bare}}(p)\widehat{\mathcal{O}}^{\tiny\mbox{bare}}(k)~,\label{eqn:OPE-bare}
\end{equation}
where $C_{\mathcal{O}}^{\tiny\mbox{bare}}$ now depends on all bare coupling constants. Since renormalized and bare operators are related by their corresponding $Z$-factors as $\widehat{O}^{{\tiny\mbox{ren}}}=Z_{\widehat{O}} \widehat{O}^{{\tiny\mbox{bare}}}$, we obtain the following relation\footnote{In case $\widehat{O}_{\I}$ (and/or $\widehat{O}_{\II}$) is a fundamental field, {e.g.}, $\phi$, we obtain $Z^{\frac{1}{2}}_{\phi}$ instead of $Z_{\widehat{O}_{\I}}$ in the numerator of the ratio \cite{Huang:2025rdy}.}
\begin{equation}
C_{\mathcal{O}}^{\tiny\mbox{ren}}(p)=
\frac{Z_{\widehat{O}_{\I}}Z_{\widehat{O}_{\II}}}{Z_{\widehat{\mathcal{O}}}}C_{\mathcal{O}}^{\tiny\mbox{bare}}(p)
~\sim~\mbox{UV~finite}~.\label{eqn:C-relation}
\end{equation}
The Wilson coefficients $C_{\mathcal{O}}^{\tiny\mbox{ren}}(p)$ of renormalized operators are UV finite, which implies that the product of the ratio of $Z$-factors and the bare Wilson coefficients $C_{\mathcal{O}}^{\tiny\mbox{bare}}(p)$ must also be UV finite. The UV finiteness conditions generate a set of algebraic equations by requiring the cancellation of $\epsilon$ poles in the RHS of eqn.(\ref{eqn:C-relation}), and the ratio of $Z$-factors can be determined once $C_{\mathcal{O}}^{\tiny\mbox{bare}}(p)$ has been computed. 

For an operator $\widehat{\mathcal{O}}^{\tiny\mbox{bare}}$ composed of $n$ fields, the bare Wilson coefficient $C_{\mathcal{O}}^{\tiny\mbox{bare}}(p)$ can be computed via the large momentum expansion of bare correlation functions as
\begin{equation}
C_{\mathcal{O}}^{\tiny\mbox{bare}}(p)=\Big\langle \widehat{O}_{\II}^{\tiny\mbox{bare}}(-\sum k_i-p)\widehat{O}_{\I}^{\tiny\mbox{bare}}(p)\varphi_0(k_1)\varphi_0(k_2)\cdots \varphi_0(k_n)\Big\rangle \Big|_{k_i\to 0~,~i=1,\ldots,n}~.\label{eqn:two-point-Wilson}
\end{equation}
As demonstrated in \cite{Huang:2024hsn}, this corresponds to factorizing the original Feynman diagram into two sub-graphs, one containing only hard external fields and the other containing only soft external fields, achieved by cutting $n$ internal and/or external lines. The bare Wilson coefficient receives contributions exclusively from hard sub-graphs, which are graphs involving two hard momenta and all other soft momenta. In momentum space, all soft momenta are set to zero, allowing $C_{\mathcal{O}}^{\tiny\mbox{bare}}(p)$ to be computed from two-point integrals. Once these two-point integrals are evaluated in the dimensional regularization scheme with a series expansion to the appropriate $\epsilon$ orders, the UV divergences contributing to the bare Wilson coefficient can be computed by summation over all two-point integrals. The UV finiteness conditions derived from eqn.(\ref{eqn:C-relation}) then fully constrain the $Z$-factors. Ultimately, beta functions and anomalous dimensions can be computed from these $Z$-factors using their respective definitions. In this algorithm, two-point integrals are systematically encoded as Wilson coefficients, ensuring its broad applicability to generic composite operators.

%%%%%%%%%%%%%%%%
\section{Constructing loop integrands}
\label{sec:integrand}
%%%%%%%%%%%%%%%%%%%

The bare Wilson coefficients are computed from their respective correlation functions. However, it is well known that computing multi-loop correlation functions is highly non-trivial, even at the integrand level. Here, we propose a method, which is abbreviated as the {\sl primitive diagram method}, for constructing loop integrands of generic field theories, with the aim of enhancing computational efficiency\footnote{As verified in explicit computations for scalar-QED theory, this method indeed exhibits better performance than the Feynman diagram method. For example, we have computed the four-scalar three-loop integrand using both methods. This integrand receives contributions from about 13 thousands Feynman graphs. The evaluation of the primitive diagram method is several times faster than that of the Feynman diagram method, and its memory management is also much superior.}. In this section, we describe the general setup and implementation of this method.

%%%%%%%%%%%%%%%%
\subsection{The primitive topology and primitive diagram}
%\label{subsec:integrand-primitive}
%%%%%%%%%%%%%%%%%%%

To begin with, let us recall the skeleton expansion technique for correlation functions. In quantum field theory, the skeleton expansion is a standard textbook technique for constructing one-Particle-Reducible (1PR) diagrams \cite{Srednicki_2007}. % Skeleton expansion described in M.Srednicki QFT book Part01 Sec19 
In this technique, Feynman diagrams are reorganized into effective tree graphs consisting of {\sl exact propagators} and {\sl exact vertices}. The exact propagator corresponds to a two-point insertion of sub-graphs. The exact vertex is a one-Particle-Irreducible (1PI) effective vertex, denoted by the vertex function $i{\bf V}(p_1,\ldots,p_n)$ for $n\geq 3$, and an $n$-point exact vertex encapsulates the contributions of all 1PI $n$-point tree and loop graphs. Note that, in principle, there are infinitely many exact vertices with arbitrary $n$, thus the effective tree graphs of a correlation function in the skeleton expansion must be constructed from all possible combinations of exact propagators and non-physical exact vertices. Compared with the traditional Feynman diagram method, the skeleton expansion technique packs all loop structures within exact propagators and vertices, such that only tree graphs need to be generated. Furthermore, once the exact propagators and vertex functions have been evaluated, they can be universally inserted into any effective tree graphs, eliminating the need for redundant computations.

\begin{figure}
  \centering
    \begin{tikzpicture}
%    \draw [help lines, step=0.5] (0,0) grid (16,5);
% 
% general primitive topology
    \node [] at (0,0) {
    \begin{tikzpicture}
      \draw [thick] (0,0)--(-1,0);
      \draw [thick] (0,0)--(-0.5,0.83);
      \draw [thick] (0,0) to [out=90,in=180] (1,0.5);
      \draw [thick] (0,0) to [out=-90,in=180] (1,-0.5);
      \draw [thick] (1.5,0)--(3,0) (1.5,0)--(2.75,0.83) (1.5,0)--(2.75,-0.83);
      \draw [red] (0,0) circle [radius=0.1];
      \draw [thick,gray,fill=lightgray] (1.5,0) circle [radius=0.75];
      \draw [fill=black] (0.5,0.2) circle [radius=0.03] (0.5,0) circle [radius=0.03] (0.5,-0.2) circle [radius=0.03] (-0.5,0.2) circle [radius=0.03] (-0.4,0.35) circle [radius=0.03] (2.6,0.2) circle [radius=0.03] (2.55,0.4) circle [radius=0.03] (2.6,-0.2) circle [radius=0.03] (2.55,-0.4) circle [radius=0.03];
      \node [left] at (-1,0) {$p_1$};
      \node [left] at (-0.5,0.83) {$p_i$};
      \node [right] at (2.75,0.83) {$p_{i+1}$};
      \node [right] at (2.75,-0.83) {$p_n$};
      \node [above] at (0.5,0.5) {$\ell_{i+1}$};
      \node [below] at (0.5,-0.5) {$\ell_j$};
      \node [] at (1.5,0) {{\tiny ($n$+$j$-2$i$)-pt}};
%      \node [] at (1.5,-0.1) {{\tiny -point}};
%      \node [] at (1,-1.5) {\footnotesize (a)};
    \end{tikzpicture}
    };
    \end{tikzpicture}
  \caption{Graph decomposition of the $n$-point correlation function. The $n$-point topology can be decomposed into a physical vertex and an $(n+j-2i)$-point correlation function. The latter admits a skeleton expansion into effective tree graphs. Different graph decompositions are identified by values $(i,j)$. For a given $j$-point physical vertex, $i$ takes integer values satisfying $1 \leq i \leq j-2$. The red circle marks the physical vertex, and the gray block represents a lower-point correlation function.}\label{fig:general-PT}
\end{figure}
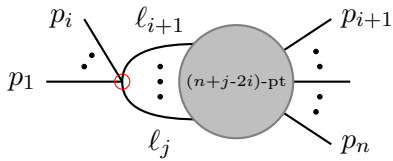

In order to provide a complete graph construction for 1PI correlation functions, let us start by considering the graph decomposition of a generic Feynman diagram, focusing on the topology of an $n$-point correlation function. If we  start from an arbitrary reference external leg (labeled by $p_1$) and traverse the line, we first encounter an interaction vertex. Assuming this is a $j$-point physical vertex, we designate it as the reference vertex. We further assume that $(i-1)$ lines from this vertex connect to other external legs, while $(j-i)$ lines connect to other interaction vertices. Here, $i$ could be any integer greater than or equal to 1, and we also require $j-i\geq 2$ to include only 1PI graphs. Within this setup, Feynman diagrams can be organized by the values of $j$ and $i$. Diagrams with the same configuration $(i,j)$ share the same graph decomposition, featuring a $j$-point physical vertex (tracked by $p_1$) connected to $(n+j-2i)$-point sub-graphs via $(j-i)$ propagators, as illustrated in Fig.(\ref{fig:general-PT}). The summation of all possible $(n+j-2i)$-point sub-graphs contributes to the $(n+j-2i)$-point correlation function, and we can convert these into effective tree graphs via skeleton expansion. In this description, all Feynman topologies are reorganized into a new classification of graphs, each featuring a physical vertex connected to an effective tree graph in the skeleton expansion.

\begin{figure}
  \centering
    \begin{tikzpicture}
%    \draw [help lines, step=0.5] (0,1) grid (16,-5);
% 
% top left, cubic primitive topology
    \node [] at (1.85,0) {
    \begin{tikzpicture}
      \draw [thick] (0,0)--(-1,0);
%      \draw [thick] (0,0)--(-0.5,0.83);
      \draw [thick] (0,0) to [out=90,in=180] (1,0.5);
      \draw [thick] (0,0) to [out=-90,in=180] (1,-0.5);
      \draw [thick] (1.5,0)--(3,0) (1.5,0)--(2.75,0.83) (1.5,0)--(2.75,-0.83);
      \draw [red] (0,0) circle [radius=0.1];
      \draw [thick,gray,fill=lightgray] (1.5,0) circle [radius=0.75];
      \draw [fill=black] (2.6,0.2) circle [radius=0.03] (2.55,0.4) circle [radius=0.03] (2.6,-0.2) circle [radius=0.03] (2.55,-0.4) circle [radius=0.03];
      \node [left] at (-1,0) {$p_1$};
      \node [right] at (2.75,0.83) {$p_{2}$};
      \node [right] at (2.75,-0.83) {$p_n$};
      \node [] at (1.5,0) {{\tiny ($n$+1)-pt}};
    \end{tikzpicture}
    };
% top middle, quartic primitiv topology case 1
    \node [] at (7.6,0) {
    \begin{tikzpicture}
      \draw [thick] (0,0)--(-1,0);
      \draw [thick] (0,0)--(-0.5,0.83);
      \draw [thick] (0,0) to [out=90,in=180] (1,0.5);
      \draw [thick] (0,0) to [out=-90,in=180] (1,-0.5);
      \draw [thick] (1.5,0)--(3,0) (1.5,0)--(2.75,0.83) (1.5,0)--(2.75,-0.83);
      \draw [red] (0,0) circle [radius=0.1];
      \draw [thick,gray,fill=lightgray] (1.5,0) circle [radius=0.75];
      \draw [fill=black] (2.6,0.2) circle [radius=0.03] (2.55,0.4) circle [radius=0.03] (2.6,-0.2) circle [radius=0.03] (2.55,-0.4) circle [radius=0.03];
      \node [left] at (-1,0) {$p_1$};
      \node [left] at (-0.5,0.83) {$p_2$};
      \node [right] at (2.75,0.83) {$p_{3}$};
      \node [right] at (2.75,-0.83) {$p_n$};
      \node [] at (1.5,0) {{\tiny $n$-pt}};
    \end{tikzpicture}
    };
% top right, quartic primitive topology case 2
    \node [] at (13.35,0) {
    \begin{tikzpicture}
      \draw [thick] (0,0)--(-1,0);
      \draw [thick] (0,0)--(1,0);
      \draw [thick] (0,0) to [out=90,in=180] (1,0.5);
      \draw [thick] (0,0) to [out=-90,in=180] (1,-0.5);
      \draw [thick] (1.5,0)--(3,0) (1.5,0)--(2.75,0.83) (1.5,0)--(2.75,-0.83);
      \draw [red] (0,0) circle [radius=0.1];
      \draw [thick,gray,fill=lightgray] (1.5,0) circle [radius=0.75];
      \draw [fill=black] (2.6,0.2) circle [radius=0.03] (2.55,0.4) circle [radius=0.03] (2.6,-0.2) circle [radius=0.03] (2.55,-0.4) circle [radius=0.03];
      \node [left] at (-1,0) {$p_1$};
      \node [right] at (2.75,0.83) {$p_{2}$};
      \node [right] at (2.75,-0.83) {$p_n$};
      \node [] at (1.5,0) {{\tiny ($n$+2)-pt}};
    \end{tikzpicture}
    };
%    \node [] at (8,-1.5) {\footnotesize (a)};
    \end{tikzpicture}
  \caption{Three types of graph decompositions for topologies of $n$-point correlation functions in scalar-QED theory, corresponding to the pairs $(i,j)=(1,3)$, $(i,j)=(2,4)$ and $(i,j)=(1,4)$ respectively. The red circle marks the physical vertex and $p_1$ denotes the reference external leg.}\label{fig:sQED-PT}
\end{figure}
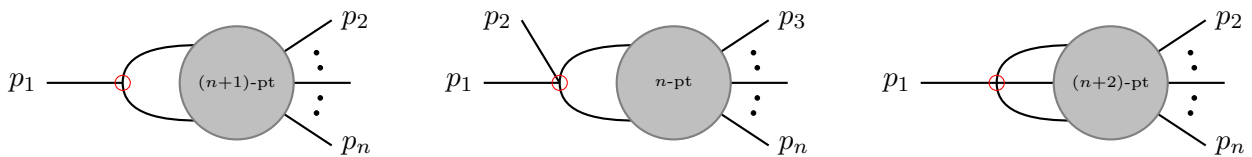

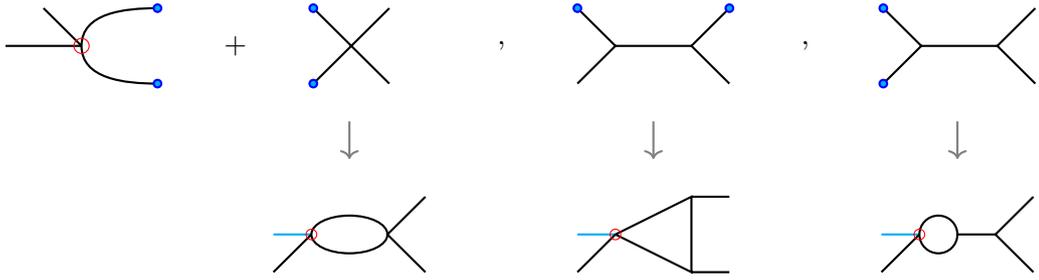
\begin{figure}
\centering
\begin{tikzpicture}
%  \draw [help lines, step=0.5] (0,0) grid (13,4);
%
% top no.1
  \node [] at (0.5,3) {
  \begin{tikzpicture}
      \draw [thick] (0,0)--(1,0) (0.5,0.5)--(1,0);
      \draw [thick] (1,0) to [out=90,in=180] (2,0.5);
      \draw [thick] (1,0) to [out=-90,in=180] (2,-0.5);
      \draw [red] (1,0) circle [radius=0.1];
      \draw [thick, blue, fill=cyan] (2,-0.5) circle [radius=0.05] (2,0.5) circle [radius=0.05];
  \end{tikzpicture}
  };
% top no.2
  \node [] at (4,3) {
  \begin{tikzpicture}
      \draw [thick] (-0.5,-0.5)--(0.5,0.5) (0.5,-0.5)--(-0.5,0.5);
      \draw [thick, blue, fill=cyan] (-0.5,-0.5) circle [radius=0.05] (-0.5,0.5) circle [radius=0.05];
  \end{tikzpicture}
  };
% top no.3
  \node [] at (8,3.025) {
  \begin{tikzpicture}
      \draw [thick] (0,0.5)--(0.5,0)--(0,-0.5) (0.5,0)--(1.5,0) (2,0.5)--(1.5,0)--(2,-0.5);
      \draw [thick, blue, fill=cyan] (0,0.5) circle [radius=0.05] (2,0.5) circle [radius=0.05];
  \end{tikzpicture}
  };
% top no.4
  \node [] at (12,3) {
  \begin{tikzpicture}
      \draw [thick] (0,0.5)--(0.5,0)--(0,-0.5) (0.5,0)--(1.5,0) (2,0.5)--(1.5,0)--(2,-0.5);
      \draw [thick, blue, fill=cyan] (0,0.5) circle [radius=0.05] (0,-0.5) circle [radius=0.05];
  \end{tikzpicture}
  };
  \node [] at (2.5,3) {+};
  \node [] at (6,3) {,};
  \node [] at (10,3) {,};
  \draw [thick, gray, ->] (4,2)--(4,1.5);
  \draw [thick, gray, ->] (8,2)--(8,1.5);
  \draw [thick, gray, ->] (12,2)--(12,1.5);  
% bottom no.1
  \node [] at (4,0.5) {
  \begin{tikzpicture}
      \draw [thick, cyan] (0,0)--(0.5,0); 
      \draw [thick] (0,-0.5)--(0.5,0) (2,0.5)--(1.5,0)--(2,-0.5);
      \draw [thick] (1,0) ellipse [x radius=0.5,y radius=0.25];
      \draw [red] (0.5,0) circle [radius=0.07];
  \end{tikzpicture}
  };
% bottom no.2
  \node [] at (8,0.5) {
  \begin{tikzpicture}
      \draw [thick, cyan] (0,0)--(0.5,0); 
      \draw [thick] (0,-0.5)--(0.5,0) (0.5,0)--(1.5,0.5)--(1.5,-0.5)--(0.5,0) (1.5,0.5)--(2,0.5) (1.5,-0.5)--(2,-0.5);
      \draw [red] (0.5,0) circle [radius=0.07];
  \end{tikzpicture}
  };
% bottom no.3
  \node [] at (12,0.5) {
  \begin{tikzpicture}
      \draw [thick, cyan] (0,0)--(0.5,0); 
      \draw [thick] (0,-0.5)--(0.5,0) (1,0)--(1.5,0) (2,0.5)--(1.5,0)--(2,-0.5);
      \draw [thick] (0.75,0) circle [radius=0.25];
      \draw [red] (0.5,0) circle [radius=0.07];
  \end{tikzpicture}
  };
\end{tikzpicture}
\caption{Example of graph decomposition for a four-point correlation function with $(i,j)=(2,4)$. The physical quartic vertex is connected to the effective tree graphs of the skeleton expansion in all possible endings, which are indicated by blue dots here. Each connection reproduces a loop graph with exact propagators and exact vertices.}\label{fig:sQED-PT-i2j4}
\end{figure}

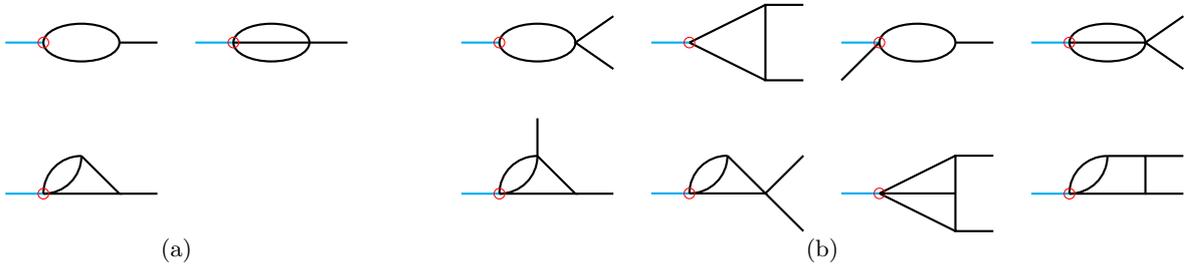
\begin{figure}
  \centering
    \begin{tikzpicture}
%    \draw [help lines, step=0.5] (0,1) grid (16,-5);
%  Middle, 1st: 2-pt cubic, 
  \node [] at (1,-2.5) {
  \begin{tikzpicture}
    \draw [cyan, thick] (0,0)--(0.5,0);
    \draw [thick] (1,0) ellipse [x radius=0.5, y radius=0.25];
    \draw [red] (0.5,0) circle [radius=0.07];
    \draw [thick] (1.5,0)--(2,0);
   \end{tikzpicture}
  };
%  Middle, 2nd: 2-pt quartic case 2
  \node [] at (3.5,-2.5) {
  \begin{tikzpicture}
    \draw [cyan, thick] (0,0)--(0.5,0);
    \draw [thick] (0.5,0)--(1.5,0);
    \draw [thick] (1,0) ellipse [x radius=0.5, y radius=0.25];
    \draw [red] (0.5,0) circle [radius=0.07];
    \draw [thick] (1.5,0)--(2,0);
   \end{tikzpicture}
  };
% Bottom, 1st: 2-pt quartic case 2
  \node [] at (1,-4.285) {
  \begin{tikzpicture}
    \draw [cyan, thick] (0,0)--(0.5,0);
    \draw [thick] (0.5,0)--(1.5,0)--(1,0.5);
    \draw [thick] (0.5,0) to [out=90,in=180] (1,0.5);
    \draw [thick] (1,0.5) to [out=270,in=0] (0.5,0);
    \draw [red] (0.5,0) circle [radius=0.07];
    \draw [thick] (2,0)--(1.5,0);
   \end{tikzpicture}
  };
% Middle, 3rd: 3-pt cubic
  \node [] at (7,-2.5) {
  \begin{tikzpicture}
    \draw [cyan, thick] (0,0)--(0.5,0);
    \draw [thick] (1,0) ellipse [x radius=0.5, y radius=0.25];
    \draw [red] (0.5,0) circle [radius=0.07];
    \draw [thick] (2,0.35)--(1.5,0)--(2,-0.35);
   \end{tikzpicture}
  };
% Middle, 4th: 3-pt cubic
  \node [] at (9.5,-2.5) {
  \begin{tikzpicture}
    \draw [cyan, thick] (0,0)--(0.5,0);
    \draw [thick] (0.5,0)--(1.5,0.5) (0.5,0)--(1.5,-0.5);
    \draw [red] (0.5,0) circle [radius=0.07];
    \draw [thick] (2,0.5)--(1.5,0.5)--(1.5,-0.5)--(2,-0.5);
   \end{tikzpicture}
  };
% Middle, 5th: 3-pt quarti case 1
  \node [] at (12,-2.625) {
  \begin{tikzpicture}
    \draw [cyan, thick] (0,0)--(0.5,0);
    \draw [thick] (0,-0.5)--(0.5,0);
    \draw [thick] (1,0) ellipse [x radius=0.5, y radius=0.25];
    \draw [red] (0.5,0) circle [radius=0.07];
    \draw [thick] (2,0)--(1.5,0);
  \end{tikzpicture}
  };
%  Middle 6th: 3-pt quartic case 2
  \node [] at (14.5,-2.5) {
  \begin{tikzpicture}
    \draw [cyan, thick] (0,0)--(0.5,0);
    \draw [thick] (0.5,0)--(1.5,0);
    \draw [thick] (1,0) ellipse [x radius=0.5, y radius=0.25];
    \draw [red] (0.5,0) circle [radius=0.07];
    \draw [thick] (2,0.35)--(1.5,0)--(2,-0.35);
   \end{tikzpicture}
  };
  % Bottom 2nd: 3-pt quartic case 2
  \node [] at (7,-4.035) {
  \begin{tikzpicture}
    \draw [cyan, thick] (0,0)--(0.5,0);
    \draw [thick] (0.5,0)--(1.5,0)--(1,0.5);
    \draw [thick] (0.5,0) to [out=90,in=180] (1,0.5);
    \draw [thick] (1,0.5) to [out=270,in=0] (0.5,0);
    \draw [red] (0.5,0) circle [radius=0.07];
    \draw [thick] (1,0.5)--(1,1);
    \draw [thick] (2,0)--(1.5,0);
   \end{tikzpicture}
  };
% Bottom 3rd: 3-pt quartic case 2
  \node [] at (9.5,-4.5) {
  \begin{tikzpicture}
    \draw [cyan, thick] (0,0)--(0.5,0);
    \draw [thick] (0.5,0)--(1.5,0)--(1,0.5);
    \draw [thick] (0.5,0) to [out=90,in=180] (1,0.5);
    \draw [thick] (1,0.5) to [out=270,in=0] (0.5,0);
    \draw [red] (0.5,0) circle [radius=0.07];
    \draw [thick] (2,0.5)--(1.5,0)--(2,-0.5);
   \end{tikzpicture}
  };
% Bottom 4th: 3-pt quartic case 2
  \node [] at (12,-4.5) {
  \begin{tikzpicture}
    \draw [cyan, thick] (0,0)--(0.5,0);
    \draw [thick] (0.5,0)--(1.5,0.5) (0.5,0)--(1.5,-0.5);
    \draw [red] (0.5,0) circle [radius=0.07];
    \draw [thick] (0.5,0)--(1.5,0);
    \draw [thick] (2,0.5)--(1.5,0.5)--(1.5,-0.5)--(2,-0.5);
   \end{tikzpicture}
  };  
% Bottom 5th: 3-pt quartic case 2
  \node [] at (14.5,-4.285) {
  \begin{tikzpicture}
    \draw [cyan, thick] (0,0)--(0.5,0);
    \draw [thick] (0.5,0)--(1.5,0)--(1.5,0.5)--(1,0.5);
    \draw [thick] (0.5,0) to [out=90,in=180] (1,0.5);
    \draw [thick] (1,0.5) to [out=270,in=0] (0.5,0);
    \draw [red] (0.5,0) circle [radius=0.07];
    \draw [thick] (2,0.5)--(1.5,0.5) (2,0)--(1.5,0);
   \end{tikzpicture}
  };
   \node [] at (2.25,-5.25) {\footnotesize (a)};
   \node [] at (10.75,-5.25) {\footnotesize (b)};
    \end{tikzpicture}
  \caption{(a) Primitive topologies of the two-point correlation function, (b) primitive topologies of the three-point correlation function in scalar-QED theory. The cyan line denotes the reference external leg, the red circle marks the reference physical vertex, and all other vertices and internal lines correspond to exact vertices and exact propagators, respectively.}\label{fig:sQED-PT-example}
\end{figure}

To make the discussion more specific, let us apply the above general graph decomposition to scalar-QED theory, in which we have $j=3$ cubic vertex and $j=4$ quartic vertex. By solving for the values of $i$ for each $j$, we obtain three types of graph decompositions, as shown in Fig.(\ref{fig:sQED-PT}).
For each type of graph decompositions, the gray block can be expressed as a collection of effective tree graphs from the skeleton expansion. For example let us consider the four-point correlation function and examine the second graph decomposition in Fig.(\ref{fig:sQED-PT}). When $n=4$, this gray block also represents a four-point correlation function. The skeleton expansion of this gray block yields two effective tree graphs, which allows for three non-isomorphic ways of connecting them to the physical vertex, as depicted in Fig.(\ref{fig:sQED-PT-i2j4}).
Note that the third connection results in a 1PR graph, which must be excluded. We refer to the remaining 1PI graphs as {\sl primitive topologies}, since all other loop topologies can be generated from them. In general, the primitive topologies of an $n$-point correlation function should be constructed from all possible graph decompositions, where the physical vertex is non-isomorphically connected to the effective tree graphs from the skeleton expansion of the gray blocks, with all 1PR results excluded. Examples of primitive topologies for two-point and three-point correlation functions in scalar-QED theory are illustrated in Fig.(\ref{fig:sQED-PT-example}). 
In a primitive topology, only the external legs and the reference vertex are physical, while all other vertices and internal lines correspond to exact vertices and exact propagators. This construction algorithm can be straightforwardly generalized to correlation functions involving an operator, where the external leg of the operator can be taken as the reference external leg and the vertex of the operator can be adopted as the reference vertex. 

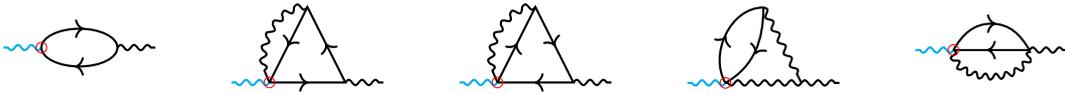
\begin{figure}
\centering
  \begin{tikzpicture}
%  \draw [help lines] (0,0) grid (15,3);
% PDVV-1
    \node [] at (0,0) {
    \begin{tikzpicture}
      \draw [thick, cyan, decorate, decoration={coil, segment length=5, aspect=0, amplitude=1}] (1,0)--(0.5,0);
      \draw [thick, decorate, decoration={coil, segment length=5, aspect=0, amplitude=1}] (2,0)--(2.5,0);
      \draw [thick] (1.5,0) ellipse [x radius=0.5,y radius=0.25];
      \draw [thick, ->] (1.55,0.25)--(1.56,0.25);
      \draw [thick, ->] (1.45,-0.25)--(1.44,-0.25);
      \draw [red] (1,0) circle [radius=0.07];
    \end{tikzpicture}
    };
% PDVV-2
    \node [] at (3,0) {
    \begin{tikzpicture}
      \draw [thick, cyan, decorate, decoration={coil, segment length=5, aspect=0, amplitude=1}] (1,0)--(0.5,0);
      \draw [thick, decorate, decoration={coil, segment length=5, aspect=0, amplitude=1}] (2,0)--(2.5,0);
      \draw [thick, decorate, decoration={coil, segment length=5, aspect=0, amplitude=1}] (1.5,1) to [out=180, in=120] (1,0);
      \draw [thick] (1,0)--(1.5,1)--(2,0)--(1,0);
      \draw [thick, ->] (1.5,0)--(1.51,0);
      \draw [thick, ->] (1.25,0.5)--(1.24,0.48);
      \draw [thick, ->] (1.75,0.5)--(1.74,0.52);
      \draw [red] (1,0) circle [radius=0.07];
    \end{tikzpicture}
    };
    \node [] at (6,0) {
    \begin{tikzpicture}
      \draw [thick, cyan, decorate, decoration={coil, segment length=5, aspect=0, amplitude=1}] (1,2)--(0.5,2);
      \draw [thick, decorate, decoration={coil, segment length=5, aspect=0, amplitude=1}] (2,2)--(2.5,2);
      \draw [thick, decorate, decoration={coil, segment length=5, aspect=0, amplitude=1}] (1.5,3) to [out=180, in=120] (1,2);
      \draw [thick] (1,2)--(1.5,3)--(2,2)--(1,2);
      \draw [thick, ->] (1.5,2)--(1.49,2);
      \draw [thick, ->] (1.25,2.5)--(1.26,2.52);
      \draw [thick, ->] (1.75,2.5)--(1.76,2.48);
      \draw [red] (1,2) circle [radius=0.07];
    \end{tikzpicture}
    };
% PDVV-3
    \node [] at (9,0) {
    \begin{tikzpicture}
      \draw [thick, cyan, decorate, decoration={coil, segment length=5, aspect=0, amplitude=1}] (1,0)--(0.5,0);
      \draw [thick, decorate, decoration={coil, segment length=5, aspect=0, amplitude=1}] (2,0)--(2.5,0);
      \draw [thick, decorate, decoration={coil, segment length=5.1, aspect=0, amplitude=1}] (1.5,1)--(2,0);
      \draw [thick, decorate, decoration={coil, segment length=5, aspect=0, amplitude=1}] (2,0)--(1,0);
      \draw [thick] (1,0) to [out=120, in=180] (1.5,1);
      \draw [thick] (1.5,1) to [out=270, in=30] (1,0);
      \draw [thick, ->] (1,0.6)--(1.01,0.62);
      \draw [thick, ->] (1.4,0.4)--(1.39,0.38);
      \draw [red] (1,0) circle [radius=0.07];
   \end{tikzpicture}
    };
 % PDVV-4
    \node [] at (12,0) {
    \begin{tikzpicture}
      \draw [thick, cyan, decorate, decoration={coil, segment length=5, aspect=0, amplitude=1}] (1,0)--(0.5,0);
      \draw [thick, decorate, decoration={coil, segment length=5, aspect=0, amplitude=1}] (2,0)--(2.5,0);
      \draw [thick, decorate, decoration={coil, segment length=4, aspect=0, amplitude=1}] (1,0) to [out=270, in=180] (1.5,-0.35) to [out=0, in=270] (2,0);
      \draw [thick] (1,0)--(2,0);
      \draw [thick] (1,0) to [out=60, in=180] (1.5,0.35) to [out=0, in=120] (2,0);
      \draw [thick, ->] (1.45,0)--(1.44,0);
      \draw [thick, ->] (1.55,0.35)--(1.56,0.35);
       \draw [red] (1,0) circle [radius=0.07];
    \end{tikzpicture}
    };
  \end{tikzpicture}
 \caption{The primitive diagrams of the two-photon correlation functions. Wavy lines denote vector fields, arrowed lines denote complex scalar fields, the cyan line marks the reference external leg, and the red circle marks the physical reference vertex.}\label{fig:sQED-PD-2V}
\end{figure}

After assigning field contents to the internal and external lines in the primitive topology, we produce the {\sl primitive diagrams}, which are reorganizations of all Feynman diagrams. The skeleton expansion technique ensures that the primitive diagrams reproduce the complete set of Feynman diagrams, and they can serve as an alternative graph description for 1PI correlation functions. We refer to this description as the {\sl primitive diagram method}. Unlike Feynman diagrams, in which the number of interaction vertices and propagators is limited, in a primitive topology, we must consider all possible non-vanishing field configurations of exact propagators and exact vertices. Although the increasing number of effective vertices and propagators may seem to complicate the computation, it dose not pose any difficulties since we only need to generate effective tree graphs, and the sub-loop structures are encoded in the effective vertices and propagators. An example of primitive diagrams for two-photon correlation functions is shown in Fig.(\ref{fig:sQED-PD-2V}).
We remark that, in these primitive diagrams, only the external legs and the reference vertex are constrained by the physical conditions specified in the Lagrangian. For instance, in the fourth primitive diagram, there is an effective three-photon vertex that is not an interaction vertex in the Lagrangian. This exact vertex has no tree-level contribution but receives non-vanishing loop corrections. Any non-vanishing vertex functions can be inserted into primitive diagrams, provided they do not violate the field assignments of the reference vertex.

%%%%%%%%%%%%%%%%
\subsection{From primitive diagram to loop integrand}
%\label{subsec:integrand-loop}
%%%%%%%%%%%%%%%%%%%

In primitive diagrams, sub-loop structures are encoded in exact propagators and exact vertices. The integrand of a correlation function at any loop order can be constructed from primitive diagrams by assigning appropriate loop orders to the exact propagators and vertices. For a generic primitive topology shown in Fig.(\ref{fig:general-PT}), its loop order is determined by the connecting lines between the physical vertex and the gray block, as well as the loop order within the gray block. The primitive topology is at least $(j-i-1)$ loop order when only tree-level contributions in the gray block are considered. Thus, when constructing an $n_\ell$-loop integrand, the loop order of the gray block should be $(n_\ell-j+i+1)$. Since $i\leq j-2$, we obtain $n_\ell-j+i+1\leq n_\ell-1$. This implies that a loop integrand is always constructed from gray blocks of lower loop order, enabling the recursive generation of loop integrands. To construct an $n_\ell$-loop integrand, the task reduces to distributing $(n_\ell-j+i+1)$-loop orders among the exact vertices and propagators. A 1-loop integrand can only be constructed from primitive topologies satisfying $i=j-2$, in which case only tree-level contributions in the gray block are permitted. In the skeleton expansion of the gray block, tree-level contributions are given by all tree graphs composed of physical propagators and vertices, thus the 1-loop integrand can be computed as products of physical propagators and vertices. Recursively, all higher-loop integrands can be constructed from lower-loop integrands, down to the 1-loop integrands that are built solely from physical propagators and interaction vertices. Notably, an $n_\nu$-point exact vertex appearing in the skeleton expansion encapsulates the contributions of all 1PI $n_\nu$-point Feynman diagrams, which by definition can also be computed via the skeleton expansion using the corresponding $n_\nu$-point primitive diagrams. An exact propagator, however, may consist of a single 1PI two-point sub-graph or a chain of multiple 1PI two-point sub-graphs. Each 1PI two-point sub-graph within the exact propagator can itself be computed via the skeleton expansion using the corresponding two-point primitive diagrams. This demonstrates that the primitive diagram method is self-contained, and all components required for loop integrand construction can be generated from known primitive diagrams.

For scalar-QED theory, as illustrated in Fig.(\ref{fig:sQED-PT}), there are only three types of graph decompositions. The first and second primitive topologies yield integrands of at least 1-loop order, while the third produces integrands of at least 2-loop order. To construct an $n$-point $n_\ell$-loop integrand, we start from any $n$-point primitive diagram and distribute $(n_\ell-1)$ or $(n_\ell-2)$ loop orders respectively among the exact propagators and exact vertices in the effective tree graphs from the skeleton expansion of the gray block, as well as the connecting lines between the reference vertex and the gray block. Ultimately, a loop integrand is decomposed into products of exact vertex functions and 1PI two-point functions, and we sum over all possible distributions of loop orders among these components to reconstruct the desired $n_\ell$-loop integrand. We define the 1PI two-loop functions and vertex functions of specific loop orders as {\sl building blocks}, as all primitive diagrams are constructed using them. These building blocks are computed once and saved for subsequent use. For scalar-QED theory, the building blocks include two-photon functions and scalar-pair functions of any loop order (physical propagators are treated as zero-th order two-point functions in this framework), as well as vertex functions of any loop order of the form 
\begin{equation}
i{\bf V}(\phi_{1},\bar{\phi}_1,\phi_2,\bar{\phi}_2,\ldots,\phi_i,\bar{\phi}_i,A^\mu_{1},A^\mu_{2},\ldots,A^\mu_j)~~~\mbox{for}~~i,j\geq 0~~,~~2i+j\geq 3~. 
\end{equation}
The physical interaction vertices $(\partial\phi)\bar{\phi}A$, $\phi\bar{\phi}A^2$ and $\phi^2\bar{\phi}^2$ are treated as zero-th order three-point and four-point exact vertices, and vertex functions with more than four points have no tree-level contributions.

We now describe the algorithm for distributing loop orders in a primitive diagram. As an illustration, consider the distribution of  $n'_\ell$ loops. First, we compile a list of all exact propagators and exact vertices present in the gray block, as well as the connecting lines between the reference vertex and the gray block. Let $n_e$ denote the total number of these components. We then generate all possible integer partitions of $n'_\ell$ where the number of factors (parts) is $\leq n_e$. For each partition, we assign the factors to the exact propagators and vertices in all possible permutations. If the number of factors in a partition is less than $n_e$, we complete the assignment by appending zeros to the partition. Special care must be taken with exact vertices that have no tree-level contributions. These vertices must be assigned a non-zero factor from the integer partition. An additional  consideration applies to the integer factor assigned to an exact propagator. Since an exact propagator may consist of one or more insertions of 1PI two-point functions, if the assigned integer factor is greater than one, we further perform an integer partition of this factor. The number of parts in the secondary partition indicates the number of 1PI two-point functions within the exact propagator. Again, for each secondary partition, we assign the resulting factors to the exact propagator in all possible permutations.

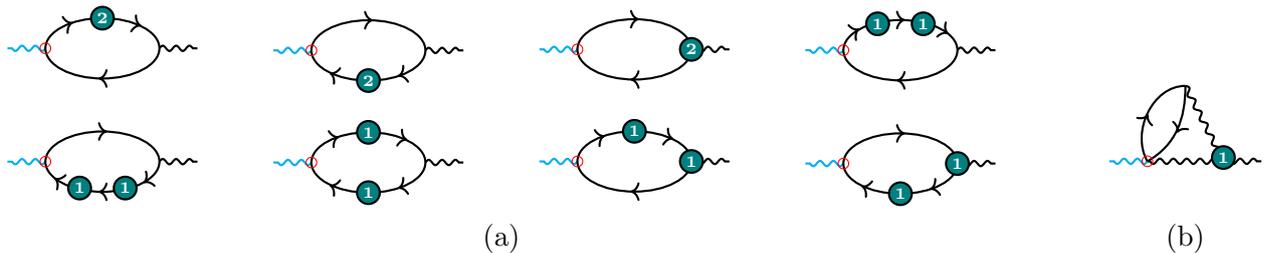
\begin{figure}
\centering
  \begin{tikzpicture}
%  \draw [help lines, step=0.5] (0,0) grid (16.5,2.5);
% top no.1
    \node [] at (1.25,2.05) {
    \begin{tikzpicture}
      \draw [thick, cyan, decorate, decoration={coil, segment length=5, aspect=0, amplitude=1}] (0.5,0)--(0,0);
      \draw [thick, decorate, decoration={coil, segment length=5, aspect=0, amplitude=1}] (2,0)--(2.5,0);
      \draw [thick] (1.25,0) ellipse [x radius=0.75,y radius=0.4];
      \draw [red] (0.5,0) circle [radius=0.07];
      % top line
      \draw [thick,fill=teal] (1.25,0.4) circle [radius=0.15];
      \node [white] at (1.25,0.4) {{\tiny \bf 2}};
      \draw [thick, ->] (1.75,0.303)--(1.76,0.3);
      \draw [thick, ->] (0.85,0.337)--(0.86,0.34);
      % bottom line
      \draw [thick, ->] (1.21,-0.4)--(1.20,-0.4);
    \end{tikzpicture}
    };
% top no.2
    \node [] at (4.75,1.95) {
    \begin{tikzpicture}
      \draw [thick, cyan, decorate, decoration={coil, segment length=5, aspect=0, amplitude=1}] (0.5,0)--(0,0);
      \draw [thick, decorate, decoration={coil, segment length=5, aspect=0, amplitude=1}] (2,0)--(2.5,0);
      \draw [thick] (1.25,0) ellipse [x radius=0.75,y radius=0.4];
      \draw [red] (0.5,0) circle [radius=0.07];
      % top line
      \draw [thick, ->] (1.3,0.4)--(1.31,0.4);
      % bottom line
      \draw [thick,fill=teal] (1.25,-0.4) circle [radius=0.15];
      \node [white] at (1.25,-0.4) {{\tiny \bf 2}};
      \draw [thick, ->] (1.65,-0.337)--(1.64,-0.34);
      \draw [thick, ->] (0.75,-0.304)--(0.74,-0.3);
    \end{tikzpicture}
    };
% top no.3
    \node [] at (8.25,2) {
    \begin{tikzpicture}
      \draw [thick, cyan, decorate, decoration={coil, segment length=5, aspect=0, amplitude=1}] (0.5,0)--(0,0);
      \draw [thick, decorate, decoration={coil, segment length=5, aspect=0, amplitude=1}] (2,0)--(2.5,0);
      \draw [thick] (1.25,0) ellipse [x radius=0.75,y radius=0.4];
      \draw [red] (0.5,0) circle [radius=0.07];
      % top line
      \draw [thick, ->] (1.3,0.4)--(1.31,0.4);
      % bottom line
      \draw [thick, ->] (1.21,-0.4)--(1.2,-0.4);
      % vertex
      \draw [thick,fill=teal] (2,0) circle [radius=0.15];
      \node [white] at (2,0) {{\tiny \bf 2}};
    \end{tikzpicture}
    };
% top no.4
    \node [] at (11.75,2) {
    \begin{tikzpicture}
      \draw [thick, cyan, decorate, decoration={coil, segment length=5, aspect=0, amplitude=1}] (0.5,0)--(0,0);
      \draw [thick, decorate, decoration={coil, segment length=5, aspect=0, amplitude=1}] (2,0)--(2.5,0);
      \draw [thick] (1.25,0) ellipse [x radius=0.75,y radius=0.4];
      \draw [red] (0.5,0) circle [radius=0.07];
      % top line
      \draw [thick,fill=teal] (0.95,0.35) circle [radius=0.15];
      \node [white] at (0.95,0.35) {{\tiny \bf 1}};
      \draw [thick,fill=teal] (1.55,0.35) circle [radius=0.15];
      \node [white] at (1.55,0.35) {{\tiny \bf 1}};
      \draw [thick, ->] (1.3,0.4)--(1.31,0.4);
      \draw [thick, ->] (0.7,0.275)--(0.71,0.28);
      \draw [thick, ->] (1.87,0.226)--(1.88,0.22);
      % bottom line
      \draw [thick, ->] (1.21,-0.4)--(1.20,-0.4);
    \end{tikzpicture}
    };
% bottom no.1
    \node [] at (1.25,0.5) {
    \begin{tikzpicture}
      \draw [thick, cyan, decorate, decoration={coil, segment length=5, aspect=0, amplitude=1}] (0.5,0)--(0,0);
      \draw [thick, decorate, decoration={coil, segment length=5, aspect=0, amplitude=1}] (2,0)--(2.5,0);
      \draw [thick] (1.25,0) ellipse [x radius=0.75,y radius=0.4];
      \draw [red] (0.5,0) circle [radius=0.07];
      % top line
      \draw [thick, ->] (1.3,0.4)--(1.31,0.4);
      % bottom line
      \draw [thick,fill=teal] (0.95,-0.35) circle [radius=0.15];
      \node [white] at (0.95,-0.35) {{\tiny \bf 1}};
      \draw [thick,fill=teal] (1.55,-0.35) circle [radius=0.15];
      \node [white] at (1.55,-0.35) {{\tiny \bf 1}};
      \draw [thick, ->] (1.21,-0.4)--(1.20,-0.4);
      \draw [thick, ->] (0.63,-0.226)--(0.62,-0.22);
      \draw [thick, ->] (1.8,-0.274)--(1.79,-0.28);
    \end{tikzpicture}
    };
% bottom no.2
    \node [] at (4.75,0.5) {
    \begin{tikzpicture}
      \draw [thick, cyan, decorate, decoration={coil, segment length=5, aspect=0, amplitude=1}] (0.5,0)--(0,0);
      \draw [thick, decorate, decoration={coil, segment length=5, aspect=0, amplitude=1}] (2,0)--(2.5,0);
      \draw [thick] (1.25,0) ellipse [x radius=0.75,y radius=0.4];
      \draw [red] (0.5,0) circle [radius=0.07];
      % top line
      \draw [thick,fill=teal] (1.25,0.4) circle [radius=0.15];
      \node [white] at (1.25,0.4) {{\tiny \bf 1}};
      \draw [thick, ->] (1.75,0.303)--(1.76,0.3);
      \draw [thick, ->] (0.85,0.337)--(0.86,0.34);
      % bottom line
      \draw [thick,fill=teal] (1.25,-0.4) circle [radius=0.15];
      \node [white] at (1.25,-0.4) {{\tiny \bf 1}};
      \draw [thick, ->] (1.65,-0.337)--(1.64,-0.34);
      \draw [thick, ->] (0.75,-0.304)--(0.74,-0.3);
    \end{tikzpicture}
    };
% bottom no.3
    \node [] at (8.25,0.55) {
    \begin{tikzpicture}
      \draw [thick, cyan, decorate, decoration={coil, segment length=5, aspect=0, amplitude=1}] (0.5,0)--(0,0);
      \draw [thick, decorate, decoration={coil, segment length=5, aspect=0, amplitude=1}] (2,0)--(2.5,0);
      \draw [thick] (1.25,0) ellipse [x radius=0.75,y radius=0.4];
      \draw [red] (0.5,0) circle [radius=0.07];
      % top line
      \draw [thick,fill=teal] (1.25,0.4) circle [radius=0.15];
      \node [white] at (1.25,0.4) {{\tiny \bf 1}};
      \draw [thick, ->] (1.75,0.303)--(1.76,0.3);
      \draw [thick, ->] (0.85,0.337)--(0.86,0.34);
      % bottom line
      \draw [thick, ->] (1.21,-0.4)--(1.20,-0.4);
      % vertex
      \draw [thick,fill=teal] (2,0) circle [radius=0.15];
      \node [white] at (2,0) {{\tiny \bf 1}};
    \end{tikzpicture}
    };
% bottom no.4
    \node [] at (11.75,0.45) {
    \begin{tikzpicture}
      \draw [thick, cyan, decorate, decoration={coil, segment length=5, aspect=0, amplitude=1}] (0.5,0)--(0,0);
      \draw [thick, decorate, decoration={coil, segment length=5, aspect=0, amplitude=1}] (2,0)--(2.5,0);
      \draw [thick] (1.25,0) ellipse [x radius=0.75,y radius=0.4];
      \draw [red] (0.5,0) circle [radius=0.07];
      % top line
      \draw [thick, ->] (1.3,0.4)--(1.31,0.4);
      % bottom line
      \draw [thick,fill=teal] (1.25,-0.4) circle [radius=0.15];
      \node [white] at (1.25,-0.4) {{\tiny \bf 1}};
      \draw [thick, ->] (1.65,-0.337)--(1.64,-0.34);
      \draw [thick, ->] (0.75,-0.304)--(0.74,-0.3);
      % vertex
      \draw [thick,fill=teal] (2,0) circle [radius=0.15];
      \node [white] at (2,0) {{\tiny \bf 1}};
    \end{tikzpicture}
    };
% bottom no.5
    \node [] at (15.5,0.95) {
    \begin{tikzpicture}
      \draw [thick, cyan, decorate, decoration={coil, segment length=5, aspect=0, amplitude=1}] (1,0)--(0.5,0);
      \draw [thick, decorate, decoration={coil, segment length=5, aspect=0, amplitude=1}] (2,0)--(2.5,0);
      \draw [thick, decorate, decoration={coil, segment length=5.1, aspect=0, amplitude=1}] (1.5,1)--(2,0);
      \draw [thick, decorate, decoration={coil, segment length=5, aspect=0, amplitude=1}] (2,0)--(1,0);
      \draw [thick] (1,0) to [out=120, in=180] (1.5,1);
      \draw [thick] (1.5,1) to [out=270, in=30] (1,0);
      \draw [thick, ->] (1,0.6)--(1.01,0.62);
      \draw [thick, ->] (1.4,0.4)--(1.39,0.38);
      \draw [red] (1,0) circle [radius=0.07];
      \draw [thick,fill=teal] (2,0.05) circle [radius=0.15];
      \node [white] at (2,0.05) {{\tiny \bf 1}};
   \end{tikzpicture}
    };
    \node [] at (6.5,-0.5) {(a)};
    \node [] at (15.5,-0.5) {(b)};
  \end{tikzpicture}
 \caption{Examples of loop distribution in primitive diagrams for the target 3-loop two-photon scalar-QED integrand. The green blocks labeled with integer numbers denote the corresponding loop-order building blocks, while the zero-th order building blocks or exact propagators are represented directly by physical propagators and interaction vertices. Wavy lines denote vector fields, arrowed lines denote complex scalar fields, the cyan line marks the reference external leg, and the red circle marks the reference vertex.}\label{fig:sQED-PD-2VL3}
\end{figure}

We demonstrate this loop distribution algorithm using the example of a two-photon correlation function, where the  target integrand is of 3-loop order. Consider the first primitive diagram in Fig.(\ref{fig:sQED-PD-2V}). It contains two exact propagators and one exact vertex, meaning we need to distribute $n'_\ell=n_\ell-1=2$ loops among these three components. The integer 2 has two distinct partitions, 2 and $1+1$. This yields the following possible assignments: $(2,0,0)$, $(0,2,0)$, $(0,0,2)$, $(1,1,0)$, $(1,0,1)$, $(0,1,1)$. Suppose the first two positions in our component list correspond to the exact propagators. When an exact propagator is assigned a factor of 2, this factor can be further partitioned into $2$ and $1+1$, corresponding to either a single 2-loop two-point insertion or two 1-loop two-point insertions within the exact propagator, respectively. By combining these secondary partitions with the initial assignments, we obtain all possible loop distributions as
\begin{equation} 
(2,0,0)~~,~~(1+1,0,0)~~,~~(0,2,0)~~,~~(0,1+1,0)~~,~~(0,0,2)~~,~~(1,1,0)~~,~~(1,0,1)~~,~~(0,1,1)~.\nonumber
\end{equation}
In total, we obtain 8 distinct loop distributions for the first primitive diagram, as illustrated in Fig.(\ref{fig:sQED-PD-2VL3}.a). Each distribution corresponds to a term computed as the product of tree-level, 1-loop and 2-loop building blocks. To obtain the complete loop integrand, we also sum the contributions from the other four primitive diagrams in Fig.(\ref{fig:sQED-PD-2V}), following the same procedure. We further examine the fourth primitive diagram, which contains four exact propagators and one exact vertex. For a 3-loop target integrand, this requires distributing  $n'_\ell=1$ loop among these five components. Notably, the exact vertex in this diagram is a three-photon vertex, which has no tree-level contribution. Consequently, the only non-zero factor (1) must be assigned to this vertex, leading to exactly one valid loop distribution for the fourth primitive diagram, as illustrate in Fig.(\ref{fig:sQED-PD-2VL3}.b). 

All the procedures described above, including the generation of primitive topologies and diagrams, the distributions of loop orders, and the multiplication of building blocks to construct loop integrands, have been systematically implemented. The scalar-QED integrands up to four-loop order used in this study were computed using this method.

%%%%%%%%%%%%%%%%
\section{The anomalous dimensions}
\label{sec:result}
%%%%%%%%%%%%%%%%%%%

Based on the proposed OPE algorithm \cite{Huang:2024hsn}, we have computed the beta functions, mass and field anomalous dimensions, and the anomalous dimensions of the $\phi^Q$ operator up to the four-loop order. The computation was implemented systematically following the procedures outlined below. First, loop integrands for various correlation functions were generated using the primitive diagram method. Second, in line with the OPE algorithm, these loop integrands were converted into two-point propagator-type integrals by setting the associated soft momenta to zero. Third, the resulting two-point integrals were reduced via the IBP method with FIRE6 \cite{Smirnov:2019qkx}, and the analytical expressions of the master integrals are available in \cite{Baikov:2010hf,Lee:2011jt}. This step yields the divergent contribution from the two-point integrals. Finally, the bare Wilson coefficients were evaluated by summing 
contributions from all relevant two-point integrals in accordance with the OPE algorithm, and their $\epsilon$ expansion series were adopted to establish the UV finiteness conditions for the renormalized Wilson coefficients, following eqn.(\ref{eqn:C-relation}). The renormalization $Z$-factors were then solved from these UV finiteness conditions. With the derived $Z$-factors, the beta functions and anomalous dimensions were readily computed by definitions. In this section, we describe the OPE employed to compute the various $Z$-factors, followed by an analysis of the gauge invariance of the anomalous dimensions. We then present the full four-loop renormalization results, with particular emphasis on those for the $\phi^Q$ operator.

%%%%%%%%%%%%%%%%
\subsection{Renormalization of scalar-QED theory via OPE algorithm}
\label{subsec:result-OPE}
%%%%%%%%%%%%%%%%%%%

The field renormalization factors $Z_{\phi}$ and $Z_A$ are derived from the two-point functions $\langle \phi\bar{\phi}\rangle$ and $\langle A_{\mu}A_{\nu}\rangle$, respectively. These correlation functions are inherently two-point integrals and can be computed using any conventional methods. As is standard practice, the field anomalous dimensions are computed by definitions as 
\begin{equation}
\gamma_{A}=\frac{1}{2}\left(\frac{\partial \ln Z_{A}}{\partial e}\beta(e)+\frac{\partial \ln Z_{A}}{\partial \lambda}\beta(\lambda)\right)~~~,~~~\gamma_{\phi}=\frac{1}{2}\left(\frac{\partial \ln Z_{\phi}}{\partial e}\beta(e)+\frac{\partial \ln Z_{\phi}}{\partial \lambda}\beta(\lambda)+\frac{\partial \ln Z_{\phi}}{\partial \xi}\frac{\partial \xi}{\partial \ln\mu}\right)~.\label{eqn:compute-field-anomalous}
\end{equation}
%

%%%%%%%%%%%%%%%%%%%
\subsection*{Renormalization of the mass}
%%%%%%%%%%%%%%%%%%%

The $Z_m$-factor of the gauge-invariant operator $\widehat{O}_m:=\phi\bar{\phi}$ can be derived from the following OPE,
\begin{equation}
\widehat{O}^{\tiny\mbox{ren}}_{m}(p)\phi(k-p)~\xrightarrow{p\to\infty}~C^{\tiny\mbox{ren}}_m(p)\phi(k)~,
\end{equation}
and the renormalized and bare Wilson coefficients satisfy the relation
\begin{equation}
C^{\tiny\mbox{ren}}_m(p)=Z_mZ_{\phi}C^{\tiny\mbox{bare}}_{m}(e_0,\lambda_0;p)~.\label{eqn:C-relation-mass}
\end{equation}
The bare Wilson coefficient is computed as
\begin{equation}
C^{\tiny\mbox{bare}}_{m}(e_0,\lambda_0;p)
=\Bigl\langle \widehat{O}_m^{\tiny\mbox{bare}}(p)\phi_0(k-p)\bar{\phi}_0(-k)\Bigr\rangle\Bigr|_{k\to 0}~,
\end{equation}
which are effectively two-point integrals derived from the three-point $\widehat{O}_m\to \phi\bar{\phi}$ correlation functions by setting the momentum of the external $\bar{\phi}$ field to zero. Once these two-point integrals are evaluated via the IBP method, the $Z_m$-factor can be determined using the UV finiteness conditions of eqn.(\ref{eqn:C-relation-mass}).

%%%%%%%%%%%%%%%%%%%
\subsection*{Renormalization of the $\phi^Q$ operator}
%%%%%%%%%%%%%%%%%%%

The $Z_{\phi^Q}$-factor of the charged operator $\phi^Q$ can be derived from the following OPE,
\begin{equation}
\phi^{Q,{\tiny\mbox{ren}}}(p)\bar{\phi}(k-p)~\xrightarrow{p\to\infty}~ C^{\tiny\mbox{ren}}_Q(p)\phi^{Q-1,{\tiny\mbox{ren}}}(k)~.
\end{equation}
The renormalized and bare Wilson coefficients are connected by the ratio of $Z$-factors as
\begin{equation}\label{eqn:C-relation-phiQ}
C^{\tiny\mbox{ren}}_Q(p)=\frac{Z_{\phi^Q}Z_{\phi}^{\frac{1}{2}}}{Z_{\phi^{Q-1}}}C^{\tiny\mbox{bare}}_{Q}(e_0,\lambda_0;p)~.
\end{equation}
This enables the recursive computation of $Z_{\phi^Q}$-factors for increasing values of $Q$. The bare Wilson coefficients can be obtained from the large momentum expansion of the following bare $\phi^Q\to Q\phi$ correlation functions
\begin{equation}
G(e_0,\lambda_0;p,k_1,\ldots, k_{Q-1}):=\Bigl\langle \phi^Q_0(p)\bar{\phi}_0(k-p)\bar{\phi}_0(k_1)\cdots \bar{\phi}_0(k_{Q-1})\Bigr\rangle~, 
\end{equation}
where $k=-(k_1+\cdots+ k_{Q-1})$. In the $p\to \infty$ limit, setting all soft momenta to zero yields
\begin{equation}
C^{\tiny\mbox{bare}}_{Q}(e_0,\lambda_0;p)=\Bigl\langle \phi^Q_0(p)\bar{\phi}_0(k-p)\bar{\phi}_0(k_1)\cdots \bar{\phi}_0(k_{Q-1})\Bigr\rangle\Bigr|_{k_i\to 0}:=\Bigl\langle \phi^Q_0(p)\bar{\phi}_0(-p)\bar{\phi}_0(0)\cdots \bar{\phi}_0(0)\Bigr\rangle~,
\end{equation}
which are two-point integrals derived from the original correlation functions by removing $Q-1$ external legs. These two-point integrals can again be evaluated via the IBP methods, allowing the bare Wilson coefficients to be computed.
The UV finiteness conditions of eqn.(\ref{eqn:C-relation-phiQ}) can then be applied to determine the ratios of $Z$-factors, leading to recursion relations for $Z_{\phi^Q}$-factors. Using the initial condition $Z_{\phi^1}=Z_{\phi}^{-\frac{1}{2}}$, $Z_{\phi^Q}$-factors can be determined for arbitrary $Q$.

%%%%%%%%%%%%%%%%%%%
\subsection*{Renormalization of the coupling constant}
%%%%%%%%%%%%%%%%%%%

The evaluation of beta functions requires the value of the bare coupling constant $\lambda_0$, which can be determined from the following OPE,
\begin{equation}
\phi(k-p)\phi(p)~\xrightarrow{p\to\infty}~ C^{\tiny\mbox{ren}}_{\phi^2}(p)\phi^{2,{\tiny\mbox{ren}}}(k)~.
\end{equation}
Again, the renormalized and bare Wilson coefficients satisfy the relation
\begin{equation}\label{eqn:C-relation-lambda}
C^{\tiny\mbox{ren}}_{\phi^2}(p)=\frac{Z_{\phi}}{Z_{\phi^2}}C^{\tiny\mbox{bare}}_{\phi^2}(e_0,\lambda_0;p)~,
\end{equation}
and the bare Wilson coefficient is computed from the large momentum expansion of the bare four-scalar correlation function as
\begin{equation}
C^{\tiny\mbox{bare}}_{\phi^2}(e_0,\lambda_0;p)
=\Bigl\langle \phi_0(-k_1-k_2-p)\phi_0(p)\bar{\phi}_0(k_1)\bar{\phi}_0(k_2)\Bigr\rangle\Bigr|_{k_1,k_2\to 0}~,
\end{equation}
which are two-point integrals derived from the $\spaa{\phi\phi\bar{\phi}\bar{\phi}}$ correlation function by setting the momenta of the two external $\bar{\phi}$ fields to zero. Given that $Z_{\phi}$, $Z_{\phi^2}$ and $e_0$ are known, the UV finiteness of eqn.(\ref{eqn:C-relation-lambda}) determines $\lambda_0$. The beta functions $\beta(e)$, $\beta(\lambda)$ can be computed from $\frac{\partial e}{\partial \ln \mu}$ and $\frac{\partial \lambda}{\partial \ln \mu}$, respectively, where the latter two are derived by solving the following matrix equation
\begin{equation}\label{eqn:compute-beta}
\begin{pmatrix}
 \frac{1}{e}+\frac{\partial \ln \widetilde{Z}_e}{\partial e} & \frac{\partial \ln \widetilde{Z}_e}{\partial \lambda} \\
  \frac{\partial \ln \widetilde{Z}_\lambda}{\partial e} & \frac{1}{\lambda}+\frac{\partial \ln \widetilde{Z}_\lambda}{\partial \lambda} \\
\end{pmatrix}
\begin{pmatrix}
\frac{\partial e}{\partial\ln \mu}\\
\frac{\partial \lambda}{\partial \ln \mu}\\
\end{pmatrix}=\begin{pmatrix}
-\epsilon\\
-2\epsilon\\
\end{pmatrix}~.
\end{equation}

\begin{figure}
\centering
  \begin{tikzpicture}
%  \draw [help lines,step=0.5] (0,0) grid (16,4);
% left
    \node [] at (3.5,1.875) {
    \begin{tikzpicture}
      \draw[thick] (0,0)--(1,0) (6,0)--(7,0);
      \draw [thick, decorate, decoration={coil, segment length=5, aspect=0, amplitude=1}] (3,0)--(1,0);
      \draw [thick, decorate, decoration={coil, segment length=5, aspect=0, amplitude=1}] (4,0)--(6,0);
      \draw [thick] (1,0)--(1,1.5) (6,0)--(6,1.5);
      \draw [thick, dashed] (1,3)--(1,1.5) (6,3)--(6,1.5);
      \draw [thick, gray, fill=lightgray, rounded corners=12] (0.75,1.25) rectangle (6.25,2.25);
%      \draw [thick, gray, fill=lightgray] (3.5,1.5) ellipse [x radius=3,y radius=0.5];
      \draw [thick, gray,fill=lightgray] (1,0) circle [radius=0.25]  (2,0) circle [radius=0.25]  (5,0) circle [radius=0.25] (6,0) circle [radius=0.25];
      \draw [fill=black] (3.5,0) circle [radius=0.05] (3.25,0) circle [radius=0.05] (3.75,0) circle [radius=0.05];
      \draw [red,thick] (0.75,0.65)--(1.25,0.85) (0.75,0.85)--(1.25,0.65);
      \draw [red,thick] (5.75,0.65)--(6.25,0.85) (5.75,0.85)--(6.25,0.65);
      \draw [thick, ->] (0.49,0)--(0.5,0);
      \draw [thick, ->] (6.51,0)--(6.5,0);
      \draw [thick, ->] (1,2.69)--(1,2.7);
      \draw [thick, ->] (6,2.69)--(6,2.7);
    \end{tikzpicture}
    };
% right
    \node [] at (12.5,1.875) {
    \begin{tikzpicture}
      \draw[thick] (0,0)--(1,0) (6,0)--(7,0);
      \draw [thick, decorate, decoration={coil, segment length=5, aspect=0, amplitude=1}] (3,0)--(1,0);
      \draw [thick, decorate, decoration={coil, segment length=5, aspect=0, amplitude=1}] (4,0)--(6,0);
      \draw [thick,dashed] (1,1)--(1,0) (6,1)--(6,0);
      \draw [thick, ->] (1,0.69)--(1,0.7);
      \draw [thick, ->] (6,0.69)--(6,0.7);
      \draw [thick, gray,fill=lightgray] (1,0) circle [radius=0.25]  (2,0) circle [radius=0.25]  (5,0) circle [radius=0.25] (6,0) circle [radius=0.25];
      \draw [fill=black] (3.5,0) circle [radius=0.05] (3.25,0) circle [radius=0.05] (3.75,0) circle [radius=0.05];
      \draw [thick, ->] (0.49,0)--(0.5,0);
      \draw [thick, ->] (6.51,0)--(6.5,0);
      \draw [thick, dashed] (2.5,3)--(2.5,2) (4.5,3)--(4.5,2);
      \draw [thick, ->] (2.5,2.69)--(2.5,2.7);
      \draw [thick, ->] (4.5,2.69)--(4.5,2.7);
      \draw [thick] (3.5,1)--(2.5,2) (3.5,1)--(4.5,2);
      \draw [fill=white, thick] (3.5,1) circle [radius=0.12];
      \draw [thick] (3.415,0.915)--(3.585,1.085) (3.415,1.085)--(3.585,0.915);
      \draw [thick, gray, fill=lightgray, rounded corners=10] (2.25,1.5) rectangle (4.75,2.25);
    \end{tikzpicture}
    };
   \draw [gray,line width=2.5,->] (7.65,2)--(8.35,2);
  \end{tikzpicture}
 \caption{A 1PI diagram of the four-scalar correlation function may generate a 1PR hard graph after applying the OPE. On the left, two lines of the original diagram are cut, which yields a soft graph in the top right, and a hard graph in the bottom right. These 1PR hard graphs contribute to the renormalization. Solid external lines represent hard scalar fields, dashed external lines represent soft scalar fields, wavy lines represent vector fields, and gray blocks represent 1PI sub-graphs. The hard graphs exhibit a chain structure.}\label{fig:4-scalar-OPE}
\end{figure}
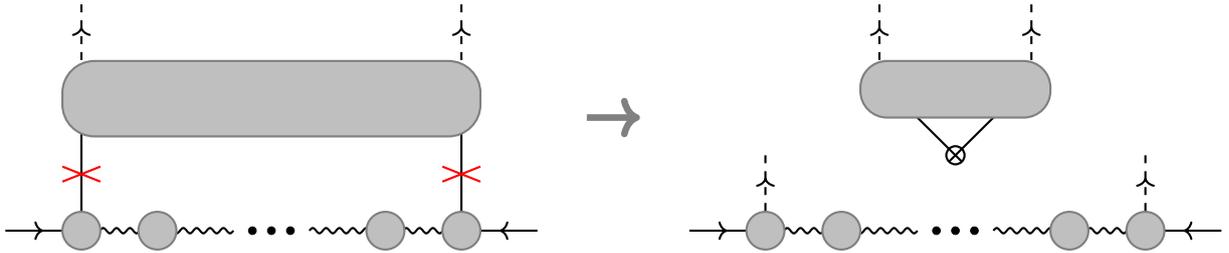

A novel aspect of this computation lies in the need to include contributions from 1PR diagrams. Conventionally, $Z$-factors are derived exclusively from 1PI correlation functions. However, in the OPE approach, the large momentum expansion of a 1PI diagram decomposes into two components, one consisting of hard diagrams carrying large momenta and the other of soft diagrams with small momenta. Crucially, although the original diagram is 1PI, the resulting hard diagrams may exhibit 1PR topologies, as shown in Fig.(\ref{fig:4-scalar-OPE}). These hard diagrams consistently take the form of chain structures. Consequently, the Wilson coefficients (corresponding to the hard diagrams) receive contributions from these chain 1PR diagrams. Using the transversality of the loop correction to photon propagator, the 1PR diagrams can be easily evaluated, as further discussed in Appendix \ref{appendix:1PR}. With contributions from 1PI and 1PR two-point integrals, the $Z_\lambda$-factor can be readily determined via the UV finiteness conditions of eqn.(\ref{eqn:C-relation-lambda}).

%

%%%%%%%%%%%%%%%%
\subsection{Gauge dependence of the anomalous dimensions}
\label{subsec:result-gauge}
%%%%%%%%%%%%%%%%%%%

In gauge theories, a critical issue is the definition of gauge-invariant correlation functions for physical quantities, as discussed, for example, in literature \cite{Kleinert:2003,Kleinert:2005sa,Irges:2017ztc}. For scalar-QED theory, the scaling dimension obtained via the semiclassical approach in \cite{Antipin:2022hfe} corresponds to the anomalous dimension of the $\phi^Q$ operator in Landau gauge. In Landau gauge, the photon propagator consists of two terms,
\begin{equation}
iD_{\mu\nu}^{\tiny\mbox{Landau}}(p)=\frac{-i}{p^2}\left(\eta_{\mu\nu}-\frac{p_{\mu}p_{\nu}}{p^2}\right)~.
\label{eqn:Landau-prop}
\end{equation}
The $\frac{1}{(p^2)^2}$ term not only significantly increases the number of terms in the correlation functions but also introduces propagators raised to higher powers, which substantially slows down the IBP reduction. In contrast, Feynman gauge is better suited for multi-loop computations, as the $\frac{1}{(p^2)^2}$ term can be neglected in the highest loop order computations where $Z_A$ can be set to unity. For lower loop computations, although the $\frac{1}{(p^2)^2}$ terms must be retained, the calculations remain more tractable compared to those at the highest loop order. Both Landau and Feynman gauges are special cases of the $R_{\xi}$-gauge, corresponding to $\xi=0$ and $\xi=1$ respectively. When including counterterm contributions, the effective photon propagator in $R_{\xi}$-gauge is given by
\begin{equation}
iD_{\mu\nu}(p)=\frac{-iZ_A^{-1}}{p^2}\left(\eta_{\mu\nu}-\frac{p_{\mu}p_{\nu}}{p^2}\right)
-i\xi\frac{p_{\mu}p_{\nu}}{(p^2)^2}~.
\label{eqn:effective-prop}
\end{equation}
In the highest loop order computations, $Z_A$ can be set to 1, and in Feynman gauge  the $\frac{1}{(p^2)^2}$ term vanishes. In this work, we evaluated the two-point correlation functions $\langle AA\rangle$ and $\langle \phi\bar{\phi}\rangle$ in $R_{\xi}$-gauge, while all other correlation functions were computed in Feynman gauge. As a result, the $Z_A$ and $Z_{\phi}$-factors were derived in $R_{\xi}$-gauge, whereas the $Z_m$, $Z_\lambda$ and $Z_{\phi^Q}$-factors were obtained in Feynman gauge. Since $Z_m$ and $\lambda_0$ are gauge-invariant and independent of $\xi$, the following discussion addresses how to determine the $\xi$-dependence of the $Z_{\phi^Q}$-factor, which is the $Z$-factor of a gauge-dependent operator.

As discussed in \cite{Zinn:2021QFT}, the following relation is all-loop exact in QED,
\begin{equation}
\frac{\partial \ln Z_{\psi}}{\partial \xi}=-\frac{e^2}{\epsilon}~.
\end{equation}
This implies that only the one-loop corrections to the anomalous dimensions exhibit dependence on the gauge parameter $\xi$, while higher loop corrections are $\xi$-independent. For scalar-QED, the analogous relation reads
\begin{equation}\label{eqn:phi-xi}
\frac{\partial \ln Z_{\phi}}{\partial \xi}=-\frac{e^2}{\epsilon}~.
\end{equation}
The same reasoning can be extended to composite operators such as $\phi^Q$, leading to the derivation of the  following relation,
\begin{equation}\label{eqn:xi-dependence}
\frac{\partial \ln Z_{\phi^Q}}{\partial \xi}=\frac{(Qe)^2}{2\epsilon}~.
\end{equation}
Using the definition $\phi^Q_R=Z_{\phi^Q}\phi^Q_0$, it follows that $Z_{\phi^1}=Z_{\phi}^{-\frac{1}{2}}$, which confirms that eqn.\eqref{eqn:phi-xi} is a special case of eqn.\eqref{eqn:xi-dependence}.

From eqn.\eqref{eqn:xi-dependence}, we can derive the expression for $\ln Z_{\phi^Q}$ in $R_{\xi}$-gauge from its counterpart in Feynman gauge as
\begin{equation}\label{eqn:z-relation-Feynman}
\ln Z_{\phi^Q}=\Bigl(\ln Z_{\phi^Q}\Bigr)\Bigr|_{\xi=1}+\frac{(Qe)^2}{2\epsilon}(\xi-1)~.
\end{equation}
The gauge parameter $\xi$ depends on the scale parameter $\mu$. It is related to the bare gauge parameter by $\xi_0=Z_A\xi$, from which we derive
\begin{equation}
\frac{\partial \xi}{\partial \ln\mu}=-\frac{\partial \ln Z_A}{\partial \ln\mu}\xi~.
\end{equation}
Thus, the anomalous dimension in $R_{\xi}$-gauge is given by
\begin{equation}
\gamma_{\phi^Q}=-\frac{\partial \ln Z_{\phi^Q}}{\partial e}\frac{\partial e}{\partial\ln\mu}
-\frac{\partial \ln Z_{\phi^Q}}{\partial \lambda}\frac{\partial \lambda}{\partial\ln\mu}
+\frac{\partial \ln Z_{\phi^Q}}{\partial \xi}\frac{\partial \ln Z_A}{\partial\ln\mu}\xi~.
\end{equation}
As evident from the explicit expression in \cite{Antipin:2022hfe}, the anomalous dimension exhibits a linear dependence on $\xi$. This linear behavior can be derived from the relation
\begin{equation}
\ln Z_{\phi^Q}=\Bigl(\ln Z_{\phi^Q}\Bigr)\Bigr|_{\xi=0}+\frac{(Qe)^2}{2\epsilon}\xi~,
\end{equation}
which is a rephrasing of eqn.\eqref{eqn:z-relation-Feynman}, combined with the observation that the combination  $e_0^2\xi_0=e^2\xi\tilde{\mu}^{2\epsilon}$ is scale independent. Consequently, the anomalous dimension in $R_{\xi}$-gauge differs from its Landau gauge counterpart by a term linear in $\xi$,
\begin{equation}
\gamma_{\phi^Q}=-\frac{\partial \ln Z_{\phi^Q}}{\partial \ln\mu}
=-\frac{\partial }{\partial \ln\mu}\Bigl(\ln Z_{\phi^Q}\Bigr)\Bigr|_{\xi=0}
-\frac{Q^2}{2\epsilon}\frac{\partial(e^2\xi)}{\partial\ln\mu}
=\gamma_{\phi^Q}^{\tiny\mbox{Landau}}
+(Qe)^2\xi~.
\end{equation}
This result is consistent with the findings in \cite{Antipin:2022hfe}. Furthermore, in \cite{Antipin:2022hfe},  Dirac's definition of a gauge-invariant non-local operator was generalized to the $\phi^Q$ operator as
\begin{equation}
\phi^Q_{nl}(x):=e^{-iQe\int d^d z J^{\mu}(z-x)A_{\mu}(x)}\phi^Q(x)~,
\end{equation}
where $J^{\mu}$ is defined as
\begin{equation}
J^{\mu}(x)=\int\frac{d^d k}{(2\pi)^d}\frac{-ik^{\mu}}{k^2}e^{ik\cdot x}\ ,
\end{equation}
and gauge invariance is guaranteed by $\partial_{\mu}J^{\mu}(x)=\delta(x)$. In the literature, this correlation function has been evaluated in coordinate space (see, {\sl e.g.}, \cite{Kleinert:2005sa}). However, evaluating it in momentum space can be more straightforward. We only need to introduce an additional Feynman rule for the vertex involving the $\phi^Q_{nl}$ operator, $n$ photons and $Q$ scalars, which is given by $(eQ)^n\prod_{i=1}^n({p_i^{\mu_i}}/{p_i^2})$. The construction of loop integrands and IBP reduction can be performed using the same approach as for the $\phi^Q$ operator. We have computed the anomalous dimensions of the $\phi^Q_{nl}$ operator up to three loops using the OPE algorithm. As expected, these anomalous dimensions coincide with those of the $\phi^Q$ operator in Landau gauge \cite{Antipin:2022hfe}.

%%%%%%%%%%%%%%%%
\subsection{Four-loop anomalous dimensions}
\label{subsec:result-AD}
%%%%%%%%%%%%%%%%%%%

Here we present various four-loop renormalization results. Renormalization computations at loop orders less than four are performed in $R_{\xi}$-gauge, allowing us to obtain the complete $\xi$-dependent $Z$-factors. While four-loop renormalization is carried out in Feynman gauge. During the four-loop computation, we set $\xi_0=1$ in the bare four-loop integrand, which eliminates the complicated double-propagator terms. This operation enables us to obtain the four-loop contributions to the $Z$-factors with $\xi=1$.

The anomalous dimension of the $\phi^Q$ operator in $R_{\xi}$-gauge up to four loops is given by
\begin{equation}
\gamma_{\phi^Q}=\sum_{n_\ell=1}^{4}\gamma_{\phi^Q}^{\tiny n_\ell\mbox{-loop}}~,\label{eqn:re-gamma-phiQ}
\end{equation}
with
\begin{eqnarray}
\gamma_{\phi^Q}^{\tiny \mbox{1-loop}}&=&e^2Q\Big(Q \xi-3\Big)+\lambda \frac{Q(Q-1)}{2}~,\\
\gamma_{\phi^Q}^{\tiny \mbox{2-loop}}&=&e^4 Q\left(Q+\frac{7}{3}\right)-2e^2 \lambda Q\Big( Q-1\Big)-\lambda^2 Q \left(\frac{Q^2}{2}-\frac{Q}{2}-\frac{1}{4}\right)~,
\end{eqnarray}
\begin{eqnarray}
\gamma_{\phi^Q}^{\tiny \mbox{3-loop}}&=&e^6 Q\left(Q^2 \Big(36 \zeta_3-4\Big)-Q \Big(108 \zeta_3-95\Big)+72 \zeta_3-\frac{3251}{54}\right)+e^4 \lambda Q \left(Q^2 \Big(-12 \zeta_3-4\Big)+\frac{29 Q}{4}+\frac{13}{4}\right)\nonumber\\
&&+e^2 \lambda^2 Q\left(Q^2 \Big(-6 \zeta_3+9\Big)+Q \Big(30 \zeta_3-31\Big)-24 \zeta_3+\frac{49}{2}\right)\nonumber\\
&&+\lambda^3 Q\left(Q^3+Q^2 \left(4 \zeta_3-\frac{17}{4}\right)-Q \left(12 \zeta_3-\frac{171}{16}\right)+8 \zeta_3-\frac{31}{4}\right)~,
\end{eqnarray}
and
\begin{eqnarray}
\gamma_{\phi^Q}^{\tiny \mbox{4-loop}}&=&e^8 Q\biggl(Q^3 \Big(-250 \zeta_5+82 \zeta_3+34\Big)+Q^2 \left(840 \zeta_5-\frac{3 \pi ^4}{5}-\frac{1600 \zeta_3}{3}-262\right)\nonumber\\
&&~~~~~~~~~~~~~~~~~~~~~+Q \left(-2200 \zeta_5-\frac{11 \pi ^4}{5}+\frac{3844 \zeta_3}{3}+\frac{69056}{81}\right)+1200 \zeta_5+2 \pi ^4-\frac{5924 \zeta_3}{9}-\frac{56189}{108}\biggr)\nonumber\\
&&+e^6 \lambda Q \biggl(Q^3 \Big(280 \zeta_5-268 \zeta_3-50\Big)+Q^2 \left(-2040 \zeta_5-\frac{4 \pi ^4}{9}+1200 \zeta_3+\frac{3994}{9}\right)\nonumber\\
&&~~~~~~~~~~~~~~~+Q \left(5120 \zeta_5+\frac{18 \pi ^4}{5}-\frac{4504 \zeta_3}{3}-\frac{2572921}{972}\right)-3520 \zeta_5-\frac{14 \pi ^4}{5}+\frac{1652 \zeta_3}{3}+\frac{2368585}{972}\biggr)\nonumber\\
&&+e^4 \lambda^2 Q\biggl(Q^3 \left(-120 \zeta_5+120 \zeta_3+\frac{99}{2}\right)+Q^2 \left(1180 \zeta_5+\frac{46 \pi ^4}{45}-575 \zeta_3-\frac{3817}{9}\right)\nonumber\\
&&~~~~~~~~~~~~~~~~+Q \left(-3400 \zeta_5-\frac{319 \pi ^4}{90}+1425 \zeta_3+\frac{18593}{12}\right)+2360 \zeta_5+\frac{106 \pi ^4}{45}-\frac{2825 \zeta_3}{3}-\frac{14575}{12}\biggr)\nonumber\\
&&+e^2 \lambda^3 Q\biggl(Q^3 \Big(40 \zeta_5-7 \zeta_3-37\Big)+Q^2 \left(-280 \zeta_5-\frac{17 \pi ^4}{30}+110 \zeta_3+\frac{387}{2}\right)\nonumber\\
&&~~~~~~~~~~~~~~~~~~~~~~~~~~~~~+Q \left(480 \zeta_5+\frac{61 \pi ^4}{30}-280 \zeta_3-\frac{16837}{48}\right)-240 \zeta_5-\frac{22 \pi ^4}{15}+176 \zeta_3+\frac{4693}{24}\biggr)\nonumber\\
&&+\lambda^4 Q\biggl(-\frac{21 Q^4}{8}+Q^3 \left( -10 \zeta_5-\frac{77 \zeta_3}{8}+\frac{71}{4}\right)+Q^2 \left(\frac{\pi ^4}{6}+\frac{47 \zeta_3}{2}-\frac{399}{8}\right)\nonumber\\
&&~~~~~~~~~~~~~~~~~~~~~~~~~~~~~~~+Q \left( 70 \zeta_5-\frac{17 \pi ^4}{30}-\frac{41 \zeta_3}{8}+\frac{235}{8}\right)-60 \zeta_5+\frac{2 \pi ^4}{5}-\frac{35 \zeta_3}{4}+\frac{509}{64}\biggr)~.
\end{eqnarray}
As expected, the gauge parameter $\xi$ appears only in the one-loop contribution. Results up to three loops have been compared with the perturbative results reported in \cite{Antipin:2022hfe}, showing exact agreements after identifying $a_e\to e^2,\lambda\to \frac{3}{2}\lambda$ in their results due to different conventions. The complete four-loop result is new. Only the leading and next-to-leading terms in the charge $Q$ expansion have been computed via semiclassical approach \cite{Antipin:2022hfe}, and they are also exactly identical to our corresponding four-loop terms after above identification.

The mass anomalous dimension for the  $\widehat{O}_m:=\phi\bar{\phi}$ operator is given by
\begin{eqnarray}
\gamma_m&=&\biggl[-6 e^2+2 \lambda\biggr]+\biggl[
\frac{86 e^4}{3}+16 e^2 \lambda-\frac{5 \lambda^2}{2}\biggr]+
\biggl[e^6 \left(-240 \zeta_3+\frac{16189}{27}\right)-e^4 \lambda \biggl(96 \zeta_3+158\biggr)\nonumber\\
&&-e^2 \lambda^2 \biggl(12 \zeta_3+7\biggr)+\frac{141 \lambda^3}{8}\biggr]+
\biggl[e^8 \left(-800 \zeta_5-\frac{212 \pi ^4}{15}-\frac{6352 \zeta_3}{9}+\frac{829}{18}\right)\nonumber\\
&&+e^6 \lambda \left(-2240 \zeta_5-\frac{16 \pi ^4}{45}+2896 \zeta_3-\frac{549469}{243}\right)+e^4 \lambda^2 \left(-520 \zeta_5+\frac{119 \pi ^4}{45}+1038 \zeta_3+\frac{9782}{9}\right)\nonumber\\
&&+e^2 \lambda^3 \left(-\frac{\pi ^4}{3}+126 \zeta_3+\frac{2275}{24}\right)-\lambda^4 \left(\frac{4 \pi ^4}{15}+\frac{25 \zeta_3}{2}+\frac{3851}{32}\right)\biggr]~.\label{eqn:re-gamma-m}
\end{eqnarray}

The four-loop beta functions and field anomalous dimensions have been computed in the literature for scalar-QED with $n$ complex scalar fields coupled to a gauge field \cite{Ihrig:2019kfv}. Using the OPE algorithm, we have reproduced these results for the $n=1$ case. Introducing $n$ complex scalars into the theory modifies the Feynman rules only by means of certain constant factors, and the complexity of their computation is nearly identical to that of the theory with a single scalar field. For the sake of presentation completeness, we also provide these results below. All relevant results have been included in the ancillary files accompanying the arXiv submission of this article. The beta functions are given by
\begin{eqnarray}
\frac{\partial e}{\partial\ln \mu}&=&-\epsilon e+\frac{e^3}{3}+4 e^5+\left[\frac{106 e^7}{9}+2 e^5 \lambda-\frac{e^3 \lambda^2}{4}\right]\nonumber\\
&&+\left[e^9\left(\frac{304 \zeta_3}{9}-\frac{16439}{243}\right)+\frac{328 e^7 \lambda}{9}-\frac{139 e^5 \lambda^2}{12}+\frac{7 e^3 \lambda^3}{8} \right]~, \label{eqn:re-beta-e}
\end{eqnarray}
and 
\begin{eqnarray}
\frac{\partial \lambda}{\partial\ln \mu}&=&-2 \epsilon \lambda+\biggl[24 e^4-12 e^2 \lambda+5 \lambda^2\biggr]+\left[
-\frac{832 e^6}{3}+\frac{316 e^4 \lambda}{3}+28 e^2 \lambda^2-15 \lambda^3\right]+\left[
e^8 \left(1152 \zeta_3-\frac{10204}{9}\right)\right.\nonumber\\
&&\left.+e^6 \lambda \left(288 \zeta_3+\frac{74714}{27}\right)-e^4 \lambda^2 \left(576 \zeta_3+\frac{1989}{2}\right)-54 e^2 \lambda^3+\lambda^4 \left(48 \zeta_3+\frac{617}{8}\right)\right]\nonumber\\
&&+\left[
e^{10} \left( 2560 \zeta_5+\frac{992 \pi ^4}{15}-\frac{97664 \zeta_3}{9}+\frac{3184112}{81}\right)+e^8 \lambda \left(-38720 \zeta_5-\frac{536 \pi ^4}{5}+\frac{60112 \zeta_3}{9}+\frac{44459}{27}\right)\right.\nonumber\\
&&+e^6 \lambda^2 \left(-21120 \zeta_5+\frac{196 \pi ^4}{15}+13480 \zeta_3-\frac{9158837}{486}\right)+e^4 \lambda^3 \left(4000 \zeta_5+\frac{64 \pi ^4}{5}+\frac{15692 \zeta_3}{3}+9325\right)\nonumber\\
&&\left.+e^2 \lambda^4 \left(-400 \zeta_5-\frac{24 \pi ^4}{5}+1110 \zeta_3+\frac{1011}{8}\right)+\lambda^5 \left(-760 \zeta_5+\frac{4 \pi ^4}{3}-514 \zeta_3-\frac{7975}{16}\right)\right]~. \label{eqn:re-beta-lambda}
\end{eqnarray}
The anomalous dimension of field $\phi$ is given by
\begin{eqnarray}
\gamma_\phi&=&e^2 \Big(\xi-3\Big)+\biggl[\frac{10 e^4}{3}+\frac{\lambda^2}{4}\biggr]+\biggl[\frac{1663 e^6}{54}+e^4 \lambda \left(-12 \zeta_3+\frac{13}{2}\right)+\frac{5 e^2 \lambda^2}{2}-\frac{5 \lambda^3}{16}\biggr]\nonumber\\
&&+\biggl[e^8 \left(-410 \zeta_5-\frac{4 \pi ^4}{5}+\frac{1546 \zeta_3}{9}+\frac{33785}{324}\right)+e^6 \lambda \left( -160 \zeta_5+\frac{16 \pi ^4}{45}-\frac{56 \zeta_3}{3}+\frac{1652}{9}\right)\nonumber\\
&&+e^4 \lambda^2 \left(20 \zeta_5-\frac{\pi ^4}{6}+\frac{85 \zeta_3}{3}-\frac{358}{9}\right)+e^2 \lambda^3 \left(-\zeta_3+\frac{61}{48}\right)+\frac{165 \lambda^4}{64}\biggr]~. \label{eqn:re-gamma-phi}
\end{eqnarray}
%

%%%%%%%%%%%%%%%%
\section{Conclusion}
\label{sec:conclusion}
%%%%%%%%%%%%%%%%%%%

In this paper, we employ the OPE algorithm proposed in \cite{Huang:2024hsn} to compute the four-loop renormalization of four-dimensional scalar-QED theory, which describes a single complex scalar field coupled to a gauge field, with particular focus on the anomalous dimensions of the fixed-charge operator $\phi^Q$. This work represents the first non-trivial validation of the OPE algorithm for multi-loop renormalization beyond pure scalar theories. We have successfully obtained the four-loop corrections to the anomalous dimensions of the $\phi^Q$ operators, as well as to the beta functions, mass and field anomalous dimensions. Our anomalous dimension of $\phi^Q$ operator agrees exactly with those corresponding terms of semiclassical computation \cite{Antipin:2022hfe}, showing non-trivial consistency. Furthermore, the gauge dependence of the anomalous dimensions of the $\phi^Q$ operators confirms the discussions on gauge-invariant operators in the semiclassical approach \cite{Antipin:2022hfe}. This successful computation further validates the efficiency and versatility of the OPE algorithm for higher-loop renormalization, and demonstrates its potential for advancing such computations to the next loop order. 

We now add a comment on the five-loop renormalization of scalar-QED. While the OPE algorithm can be straightforwardly extended to this order, and master integrals for five-loop two-point propagator-type graphs are available \cite{Georgoudis:2018olj,Georgoudis:2021onj}, the construction of loop integrands remains a major computational bottleneck. The primitive diagram method proposed in this paper provides a promising pathway for extending integrand construction to higher-loop orders, and has already demonstrated its remarkable efficacy in four-loop computations. This method makes the construction of five-loop integrands for renormalization purposes feasible in the near future. Nevertheless, IBP reduction for five-loop integrals still poses a great challenge, and the available algorithms may lack the efficiency required for five-loop scalar-QED integrals. Ongoing work focuses on advancing the primitive diagram method as well as its corresponding algorithm, and we therefore defer the five-loop renormalization to a future study.

Investigations across diverse scenarios will further uncover the potential of the OPE algorithm in renormalization computations, while simultaneously advancing the development of various traditional methods for loop integrand construction, integral reduction, integral evaluation, UV divergence extraction, and related techniques. Explorations of the OPE algorithm will primarily focus on three directions. First, extending renormalization computations to the next loop order for field theories already tested by the OPE algorithm, for example the 7-loop anomalous dimensions of the $\phi^Q$ operator in four-dimensional quartic scalar theory, the 6-loop anomalous dimensions in six-dimensional cubic scalar theory, and the aforementioned 5-loop anomalous dimensions in scalar-QED theory. Second, testing and validating the OPE algorithm in other field theories, such as the Gross-Neveu model \cite{Gracey:2016mio}, the Gross-Neveu-Yukawa model \cite{Gracey:2025aoj}, and the standard model \cite{Baikov:2016tgj,Bednyakov:2023iuj}. Third, applying the OPE algorithm to compute the anomalous dimensions of more intricate operators, including those with higher derivatives, Lorentz indices, or in scenarios involving operator mixing \cite{Derkachov:1997ch,Irges:2019bzb,Cao:2021cdt}. All such explorations require further improvements to the underlying methodologies as well as the continuous updating of computational codes. We hope that the future development of the OPE algorithm will inspire new ideas and technologies for high-precision computations in quantum field theories.

%%%%%%%%%%%%%%%%%%%%%%%
\section*{Acknowledgments}
%%%%%%%%%%%%%%%%%%%%

We would like to thank Ruizhen Huang and Wei Zhu for inspiring discussions. %RH was supported by the National Natural Science Foundation of China (NSFC) with Grant No.11805102.
This work is supported by the Science Challenge Project (No. TZ2025012), and NSAF No.U2330401.

%%%%%%%%%%%%%%%%%%%%%%%%%%
\appendix
%%%%%%%%%%%%%%%%%%%%%

%%%%%%%%%%%%%%%%%%%%%%%%%%%%%%%%%%%%%%%%%%%
\section{Contributions of 1PR diagrams in four-scalar correlation function}
\label{appendix:1PR}
%%%%%%%%%%%%%%%%%%%%%%%%%%%%%%%%%%%%%%%%%%%

The bare Wilson coefficient in the OPE of four-scalar correlation functions also receives contributions from 1PR Feynman diagrams. All contributing 1PR diagrams exhibit a chain structure, wherein two-point sub-graphs are connected in a chain via vector propagators, as already shown in Fig.(\ref{fig:4-scalar-OPE}). These diagrams can be represented schematically as follows,
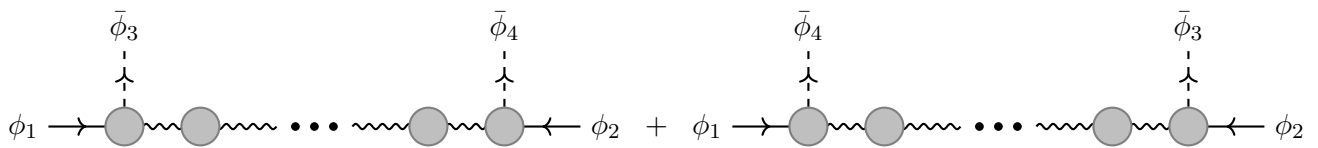
\begin{center}
  \begin{tikzpicture}
%  \draw [help lines,step=0.5] (0,0) grid (16,2);
% left
    \node [] at (3.5,0.875) {
    \begin{tikzpicture}
      \draw[thick] (0,0)--(1,0) (6,0)--(7,0);
      \draw [thick, decorate, decoration={coil, segment length=5, aspect=0, amplitude=1}] (3,0)--(1,0);
      \draw [thick, decorate, decoration={coil, segment length=5, aspect=0, amplitude=1}] (4,0)--(6,0);
      \draw [thick,dashed] (1,1)--(1,0) (6,1)--(6,0);
      \draw [thick, ->] (1,0.69)--(1,0.7);
      \draw [thick, ->] (6,0.69)--(6,0.7);
      \draw [thick, gray,fill=lightgray] (1,0) circle [radius=0.25]  (2,0) circle [radius=0.25]  (5,0) circle [radius=0.25] (6,0) circle [radius=0.25];
      \draw [fill=black] (3.5,0) circle [radius=0.05] (3.25,0) circle [radius=0.05] (3.75,0) circle [radius=0.05];
      \draw [thick, ->] (0.49,0)--(0.5,0);
      \draw [thick, ->] (6.51,0)--(6.5,0);
    \end{tikzpicture}
    };
    \node [left] at (0,0.5) {$\phi_1$}; 
    \node [right] at (7,0.5) {$\phi_2$};
    \node [above] at (1,1.5) {$\bar{\phi}_3$}; 
    \node [above] at (6,1.5) {$\bar{\phi}_4$}; 
% right
    \node [] at (12.5,0.875) {
    \begin{tikzpicture}
      \draw[thick] (0,0)--(1,0) (6,0)--(7,0);
      \draw [thick, decorate, decoration={coil, segment length=5, aspect=0, amplitude=1}] (3,0)--(1,0);
      \draw [thick, decorate, decoration={coil, segment length=5, aspect=0, amplitude=1}] (4,0)--(6,0);
      \draw [thick,dashed] (1,1)--(1,0) (6,1)--(6,0);
      \draw [thick, ->] (1,0.69)--(1,0.7);
      \draw [thick, ->] (6,0.69)--(6,0.7);
      \draw [thick, gray,fill=lightgray] (1,0) circle [radius=0.25]  (2,0) circle [radius=0.25]  (5,0) circle [radius=0.25] (6,0) circle [radius=0.25];
      \draw [fill=black] (3.5,0) circle [radius=0.05] (3.25,0) circle [radius=0.05] (3.75,0) circle [radius=0.05];
      \draw [thick, ->] (0.49,0)--(0.5,0);
      \draw [thick, ->] (6.51,0)--(6.5,0);
    \end{tikzpicture}
    };
    \node [left] at (9,0.5) {$\phi_1$}; 
    \node [right] at (16,0.5) {$\phi_2$};
    \node [above] at (10,1.5) {$\bar{\phi}_4$}; 
    \node [above] at (15,1.5) {$\bar{\phi}_3$}; 
   \node [] at (8,0.5) {+};
  \end{tikzpicture}
\end{center}
Here, the field $\phi_1$ is connected to a three-point exact vertex, while the field $\phi_2$ is connected to another three-point exact vertex. The intermediate sub-graphs are collectively described by the two-photon exact propagator $\triangle^{\mu\nu}$. On this basis, the contributions of 1PR diagrams can be expressed as
\begin{equation}
i{\bf V}_{\mu}(p_1,p_3)\triangle^{\mu\nu}(p_{13})i{\bf V}_{\nu}(p_2,p_4)
+i{\bf V}_{\mu}(p_1,p_4)\triangle^{\mu\nu}(p_{14})i{\bf V}_{\nu}(p_2,p_3)~,\label{eqn:1PR-integrand}
\end{equation}
where $p_{ij}:=p_i+p_j$, and 
\begin{equation}
i{\bf V}_{\mu}(p,p'):=\Bigl\langle \phi(p)\bar{\phi}(p')A_{\mu}(-p-p')\Bigr\rangle~.
\end{equation}
The exact propagator takes the general form as
\begin{equation}
\triangle_{\mu\nu}(p)=-i\xi\frac{p_{\mu}p_{\nu}}{(p^2)^2}
+\frac{-i}{p^2}\left(\eta_{\mu\nu}-\frac{p_{\mu}p_{\nu}}{p^2}\right)
\frac{Z_A^{-1}}{1+\frac{i}{(d-1)p^2}\Pi^{\mu}_{\ \mu}(p,g_0,Z_A \xi)}~.\label{eqn:exact-prop}
\end{equation}
Following the OPE algorithm for renormalization, we consider the leading order contributions of the large momentum expansion by setting the momenta of soft $\bar{\phi}$ fields to zero, and perform the large momentum limit in eqn.(\ref{eqn:1PR-integrand}) according to
\begin{equation}
p_1\to p~~~,~~~p_2\to -p~~~,~~~p_3\to 0~~~,~~~p_4\to 0~.
\end{equation}
In this case, the vertex functions reduce to
\begin{equation}
i{\bf V}_{\mu}(p_1,p_3)~~\to~~i{\bf V}_{\mu}(p,0):=V(p)p_\mu~~~,~~~i{\bf V}_{\nu}(p_2,p_4)~~\to~~i{\bf V}_{\nu}(-p,0):=-V(p)p_\nu~.
\end{equation}
When the Lorentz indices are contracted, the identity $p_\mu p_\nu(\eta^{\mu\nu}-p_\mu p_\nu/p^2)=0$ holds. As a result, only the first term in the exact propagator (\ref{eqn:exact-prop}) contributes, which eliminates all potential complicated computations associated with photon self-energy corrections. Thus eqn.(\ref{eqn:1PR-integrand}) evaluates to
\begin{equation}
2\biggl(V(p)p_\mu\biggr)\biggl(-i\xi\frac{p^\mu p^\nu}{(p^2)^2}\biggr)\biggl(-V(p)p_\nu\biggr)=2i\xi V^2(p)~.
\end{equation}
This simplifies the computation of 1PR diagrams to the evaluation of three-point vertex functions in the large momentum limit.

\bibliographystyle{JHEP}
\bibliography{Hbib}

\providecommand{\href}[2]{#2}\begingroup\raggedright\begin{thebibliography}{100}

\bibitem{dirac1927quantum}
P.~A.~M. Dirac, {\it {The Quantum Theory of the Emission and Absorption of
  Radiation}},  {\em Proceedings of the Royal Society A} {\bf 114} (1927),
  no.~767 243--265.

\bibitem{Zinn:2021QFT}
J.~Zinn-Justin, {\em {Quantum Field Theory and Critical Phenomena: Fifth
  Edition}}.
\newblock Oxford University Press, 04, 2021.

\bibitem{Shankar:2017zag}
R.~Shankar, {\em {Quantum Field Theory and Condensed Matter}}.
\newblock Cambridge University Press, 8, 2017.

\bibitem{Anderson:1963pc}
P.~W. Anderson, {\it {Plasmons, Gauge Invariance, and Mass}},  {\em Phys. Rev.}
  {\bf 130} (1963) 439--442.

\bibitem{Englert:1964et}
F.~Englert and R.~Brout, {\it {Broken Symmetry and the Mass of Gauge Vector
  Mesons}},  {\em Phys. Rev. Lett.} {\bf 13} (1964) 321--323.

\bibitem{Higgs:1964pj}
P.~W. Higgs, {\it {Broken Symmetries and the Masses of Gauge Bosons}},  {\em
  Phys. Rev. Lett.} {\bf 13} (1964) 508--509.

\bibitem{Guralnik:1964eu}
G.~S. Guralnik, C.~R. Hagen, and T.~W.~B. Kibble, {\it {Global Conservation
  Laws and Massless Particles}},  {\em Phys. Rev. Lett.} {\bf 13} (1964)
  585--587.

\bibitem{Ginzburg:1950sr}
V.~L. Ginzburg and L.~D. Landau, {\it {On the Theory of Superconductivity}},
  {\em Zh. Eksp. Teor. Fiz.} {\bf 20} (1950) 1064--1082.

\bibitem{Hohenberg:2015jgf}
P.~Hohenberg and A.~Krekhov, {\it {An introduction to the Ginzburg–Landau
  theory of phase transitions and nonequilibrium patterns}},  {\em Physics
  Reports} {\bf 572} (Apr., 2015) 1–42.

\bibitem{Halperin:1973jh}
B.~i. Halperin, T.~C. Lubensky, and S.-k. Ma, {\it {First order phase
  transitions in superconductors and smectic A liquid crystals}},  {\em Phys.
  Rev. Lett.} {\bf 32} (1974) 292--295.

\bibitem{Herbut:1996ut}
I.~F. Herbut and Z.~Tesanovic, {\it {Critical fluctuations in superconductors
  and the magnetic field penetration depth}},  {\em Phys. Rev. Lett.} {\bf 76}
  (1996) 4588, [\href{http://arxiv.org/abs/cond-mat/9605185}{{\tt
  cond-mat/9605185}}].

\bibitem{PhysRevLett.70.1501}
X.-G. Wen and Y.-S. Wu, {\it {Transitions between the quantum Hall states and
  insulators induced by periodic potentials}},  {\em Phys. Rev. Lett.} {\bf 70}
  (Mar, 1993) 1501--1504.

\bibitem{RevModPhys.65.851}
M.~C. Cross and P.~C. Hohenberg, {\it {Pattern formation outside of
  equilibrium}},  {\em Rev. Mod. Phys.} {\bf 65} (Jul, 1993) 851--1112.

\bibitem{Kibble:1976sj}
T.~W.~B. Kibble, {\it {Topology of Cosmic Domains and Strings}},  {\em J. Phys.
  A} {\bf 9} (1976) 1387--1398.

\bibitem{Hindmarsh:2008dw}
M.~Hindmarsh, S.~Stuckey, and N.~Bevis, {\it {Abelian Higgs Cosmic Strings:
  Small Scale Structure and Loops}},  {\em Phys. Rev. D} {\bf 79} (2009)
  123504, [\href{http://arxiv.org/abs/0812.1929}{{\tt arXiv:0812.1929}}].

\bibitem{Dufaux:2010cf}
J.-F. Dufaux, D.~G. Figueroa, and J.~Garcia-Bellido, {\it {Gravitational Waves
  from Abelian Gauge Fields and Cosmic Strings at Preheating}},  {\em Phys.
  Rev. D} {\bf 82} (2010) 083518, [\href{http://arxiv.org/abs/1006.0217}{{\tt
  arXiv:1006.0217}}].

\bibitem{Hindmarsh:2017qff}
M.~Hindmarsh, J.~Lizarraga, J.~Urrestilla, D.~Daverio, and M.~Kunz, {\it
  {Scaling from gauge and scalar radiation in Abelian Higgs string networks}},
  {\em Phys. Rev. D} {\bf 96} (2017), no.~2 023525,
  [\href{http://arxiv.org/abs/1703.06696}{{\tt arXiv:1703.06696}}].

\bibitem{Hindmarsh:2021mnl}
M.~Hindmarsh, J.~Lizarraga, A.~Urio, and J.~Urrestilla, {\it {Loop decay in
  Abelian-Higgs string networks}},  {\em Phys. Rev. D} {\bf 104} (2021), no.~4
  043519, [\href{http://arxiv.org/abs/2103.16248}{{\tt arXiv:2103.16248}}].

\bibitem{Affleck:1981bma}
I.~K. Affleck, O.~Alvarez, and N.~S. Manton, {\it {Pair Production at Strong
  Coupling in Weak External Fields}},  {\em Nucl. Phys. B} {\bf 197} (1982)
  509--519.

\bibitem{Dunne:2005sx}
G.~V. Dunne and C.~Schubert, {\it {Worldline instantons and pair production in
  inhomogeneous fields}},  {\em Phys. Rev. D} {\bf 72} (2005) 105004,
  [\href{http://arxiv.org/abs/hep-th/0507174}{{\tt hep-th/0507174}}].

\bibitem{Wilson:1971dc}
K.~G. Wilson and M.~E. Fisher, {\it {Critical exponents in 3.99 dimensions}},
  {\em Phys. Rev. Lett.} {\bf 28} (1972) 240--243.

\bibitem{PELISSETTO2002549}
A.~Pelissetto and E.~Vicari, {\it {Critical phenomena and renormalization-group
  theory}},  {\em Phys. Rept.} {\bf 368} (2002), no.~6 549--727.

\bibitem{Moshe:2003xn}
M.~Moshe and J.~Zinn-Justin, {\it {Quantum field theory in the large N limit: A
  Review}},  {\em Phys. Rept.} {\bf 385} (2003) 69--228,
  [\href{http://arxiv.org/abs/hep-th/0306133}{{\tt hep-th/0306133}}].

\bibitem{Vladimirov:1979ib}
A.~A. Vladimirov and D.~V. Shirkov, {\it {THE RENORMALIZATION GROUP AND
  ULTRAVIOLET ASYMPTOTICS}},  {\em Sov. Phys. Usp.} {\bf 22} (1979) 860--878.

\bibitem{vanDamme:1982bg}
R.~M.~J. van Damme, {\it {Most General Two Loop Counterterm for Fermion Free
  Gauge Theories With Scalar Fields}},  {\em Phys. Lett. B} {\bf 110} (1982)
  239--241.

\bibitem{PhysRevB.41.4083}
S.~Kolnberger and R.~Folk, {\it {Critical fluctuations in superconductors}},
  {\em Phys. Rev. B} {\bf 41} (Mar, 1990) 4083--4088.

\bibitem{Ihrig:2019kfv}
B.~Ihrig, N.~Zerf, P.~Marquard, I.~F. Herbut, and M.~M. Scherer, {\it {Abelian
  Higgs model at four loops, fixed-point collision and deconfined
  criticality}},  {\em Phys. Rev. B} {\bf 100} (2019), no.~13 134507,
  [\href{http://arxiv.org/abs/1907.08140}{{\tt arXiv:1907.08140}}].

\bibitem{Batkovich:2016jus}
D.~V. Batkovich, K.~G. Chetyrkin, and M.~V. Kompaniets, {\it {Six loop
  analytical calculation of the field anomalous dimension and the critical
  exponent $\eta$ in $O(n)$-symmetric $\varphi^4$ model}},  {\em Nucl. Phys. B}
  {\bf 906} (2016) 147--167, [\href{http://arxiv.org/abs/1601.01960}{{\tt
  arXiv:1601.01960}}].

\bibitem{Kompaniets:2016hct}
M.~Kompaniets and E.~Panzer, {\it {Renormalization group functions of $\phi^4$
  theory in the MS-scheme to six loops}},  {\em PoS} {\bf LL2016} (2016) 038,
  [\href{http://arxiv.org/abs/1606.09210}{{\tt arXiv:1606.09210}}].

\bibitem{Kompaniets:2017yct}
M.~V. Kompaniets and E.~Panzer, {\it {Minimally subtracted six loop
  renormalization of $O(n)$-symmetric $\phi^4$ theory and critical exponents}},
   {\em Phys. Rev. D} {\bf 96} (2017), no.~3 036016,
  [\href{http://arxiv.org/abs/1705.06483}{{\tt arXiv:1705.06483}}].

\bibitem{Schnetz:2016fhy}
O.~Schnetz, {\it {Numbers and Functions in Quantum Field Theory}},  {\em Phys.
  Rev. D} {\bf 97} (2018), no.~8 085018,
  [\href{http://arxiv.org/abs/1606.08598}{{\tt arXiv:1606.08598}}].

\bibitem{Schnetz:2025opm}
O.~Schnetz, {\it {Graphical functions with spin}},  {\em JHEP} {\bf 06} (2025)
  053, [\href{http://arxiv.org/abs/2504.05850}{{\tt arXiv:2504.05850}}].

\bibitem{Gracey:2015tta}
J.~A. Gracey, {\it {Four loop renormalization of $\phi^3$ theory in six
  dimensions}},  {\em Phys. Rev. D} {\bf 92} (2015), no.~2 025012,
  [\href{http://arxiv.org/abs/1506.03357}{{\tt arXiv:1506.03357}}].

\bibitem{Kompaniets:2021hwg}
M.~Kompaniets and A.~Pikelner, {\it {Critical exponents from five-loop scalar
  theory renormalization near six-dimensions}},  {\em Phys. Lett. B} {\bf 817}
  (2021) 136331, [\href{http://arxiv.org/abs/2101.10018}{{\tt
  arXiv:2101.10018}}].

\bibitem{Borinsky:2021jdb}
M.~Borinsky, J.~A. Gracey, M.~V. Kompaniets, and O.~Schnetz, {\it {Five-loop
  renormalization of \ensuremath{\phi}3 theory with applications to the
  Lee-Yang edge singularity and percolation theory}},  {\em Phys. Rev. D} {\bf
  103} (2021), no.~11 116024, [\href{http://arxiv.org/abs/2103.16224}{{\tt
  arXiv:2103.16224}}].

\bibitem{Schnetz:2025wtu}
O.~Schnetz, {\it {{\ensuremath{\phi}}3 theory at six loops}},  {\em Phys. Rev.
  D} {\bf 112} (2025), no.~1 016028,
  [\href{http://arxiv.org/abs/2505.15485}{{\tt arXiv:2505.15485}}].

\bibitem{DePrato:2003yd}
M.~De~Prato, A.~Pelissetto, and E.~Vicari, {\it {Third harmonic exponent in
  three-dimensional N vector models}},  {\em Phys. Rev. B} {\bf 68} (2003)
  092403, [\href{http://arxiv.org/abs/cond-mat/0302145}{{\tt
  cond-mat/0302145}}].

\bibitem{Calabrese:2004ca}
P.~Calabrese and P.~Parruccini, {\it {Harmonic crossover exponents in O(n)
  models with the pseudo-epsilon expansion approach}},  {\em Phys. Rev. B} {\bf
  71} (2005) 064416, [\href{http://arxiv.org/abs/cond-mat/0411027}{{\tt
  cond-mat/0411027}}].

\bibitem{Arias-Tamargo:2019xld}
G.~Arias-Tamargo, D.~Rodriguez-Gomez, and J.~G. Russo, {\it {The large charge
  limit of scalar field theories and the Wilson-Fisher fixed point at
  $\epsilon=0$}},  {\em JHEP} {\bf 10} (2019) 201,
  [\href{http://arxiv.org/abs/1908.11347}{{\tt arXiv:1908.11347}}].

\bibitem{Badel:2019oxl}
G.~Badel, G.~Cuomo, A.~Monin, and R.~Rattazzi, {\it {The Epsilon Expansion
  Meets Semiclassics}},  {\em JHEP} {\bf 11} (2019) 110,
  [\href{http://arxiv.org/abs/1909.01269}{{\tt arXiv:1909.01269}}].

\bibitem{Antipin:2020abu}
O.~Antipin, J.~Bersini, F.~Sannino, Z.-W. Wang, and C.~Zhang, {\it {Charging
  the $O(N)$ model}},  {\em Phys. Rev. D} {\bf 102} (2020), no.~4 045011,
  [\href{http://arxiv.org/abs/2003.13121}{{\tt arXiv:2003.13121}}].

\bibitem{Giombi:2020enj}
S.~Giombi and J.~Hyman, {\it {On the large charge sector in the critical O(N)
  model at large N}},  {\em JHEP} {\bf 09} (2021) 184,
  [\href{http://arxiv.org/abs/2011.11622}{{\tt arXiv:2011.11622}}].

\bibitem{Arias-Tamargo:2020fow}
G.~Arias-Tamargo, D.~Rodriguez-Gomez, and J.~G. Russo, {\it {On the UV
  completion of the $O(N)$ model in $6-\epsilon$ dimensions: a stable
  large-charge sector}},  {\em JHEP} {\bf 09} (2020) 064,
  [\href{http://arxiv.org/abs/2003.13772}{{\tt arXiv:2003.13772}}].

\bibitem{Antipin:2021jiw}
O.~Antipin, J.~Bersini, F.~Sannino, Z.-W. Wang, and C.~Zhang, {\it {More on the
  cubic versus quartic interaction equivalence in the $O(N)$ model}},  {\em
  Phys. Rev. D} {\bf 104} (2021) 085002,
  [\href{http://arxiv.org/abs/2107.02528}{{\tt arXiv:2107.02528}}].

\bibitem{Giombi:2022gjj}
S.~Giombi, E.~Helfenberger, and H.~Khanchandani, {\it {Long range, large
  charge, large N}},  {\em JHEP} {\bf 01} (2023) 166,
  [\href{http://arxiv.org/abs/2205.00500}{{\tt arXiv:2205.00500}}].

\bibitem{Antipin:2022naw}
O.~Antipin, J.~Bersini, and P.~Panopoulos, {\it {Yukawa interactions at large
  charge}},  {\em JHEP} {\bf 10} (2022) 183,
  [\href{http://arxiv.org/abs/2208.05839}{{\tt arXiv:2208.05839}}].

\bibitem{Antipin:2022hfe}
O.~Antipin, A.~Bednyakov, J.~Bersini, P.~Panopoulos, and A.~Pikelner, {\it
  {Gauge Invariance at Large Charge}},  {\em Phys. Rev. Lett.} {\bf 130}
  (2023), no.~2 021602, [\href{http://arxiv.org/abs/2210.10685}{{\tt
  arXiv:2210.10685}}].

\bibitem{Antipin:2023tar}
O.~Antipin, J.~Bersini, P.~Panopoulos, F.~Sannino, and Z.-W. Wang, {\it
  {Infinite order results for charged sectors of the Standard Model}},  {\em
  JHEP} {\bf 02} (2024) 168, [\href{http://arxiv.org/abs/2312.12963}{{\tt
  arXiv:2312.12963}}].

\bibitem{Jack:2021ypd}
I.~Jack and D.~R.~T. Jones, {\it {Anomalous dimensions at large charge in d=4
  O(N) theory}},  {\em Phys. Rev. D} {\bf 103} (2021), no.~8 085013,
  [\href{http://arxiv.org/abs/2101.09820}{{\tt arXiv:2101.09820}}].

\bibitem{Jin:2022nqq}
Q.~Jin and Y.~Li, {\it {Five-loop anomalous dimensions of
  \ensuremath{\phi}$^{Q}$ operators in a scalar theory with O(N) symmetry}},
  {\em JHEP} {\bf 10} (2022) 084, [\href{http://arxiv.org/abs/2205.02535}{{\tt
  arXiv:2205.02535}}].

\bibitem{Bednyakov:2022guj}
A.~Bednyakov and A.~Pikelner, {\it {Six-loop anomalous dimension of the
  \ensuremath{\phi}Q operator in the O(N) symmetric model}},  {\em Phys. Rev.
  D} {\bf 106} (2022), no.~7 076015,
  [\href{http://arxiv.org/abs/2208.04612}{{\tt arXiv:2208.04612}}].

\bibitem{Jack:2021ziq}
I.~Jack and D.~R.~T. Jones, {\it {Scaling dimensions at large charge for cubic
  \ensuremath{\phi}3 theory in six dimensions}},  {\em Phys. Rev. D} {\bf 105}
  (2022), no.~4 045021, [\href{http://arxiv.org/abs/2112.01196}{{\tt
  arXiv:2112.01196}}].

\bibitem{Huang:2024hsn}
R.~Huang, Q.~Jin, and Y.~Li, {\it {From operator product expansion to anomalous
  dimensions}},  {\em JHEP} {\bf 06} (2025) 135,
  [\href{http://arxiv.org/abs/2410.03283}{{\tt arXiv:2410.03283}}].

\bibitem{Huang:2025rdy}
R.~Huang, Q.~Jin, and Y.~Li, {\it {Five-loop anomalous dimensions of cubic
  scalar theory from operator product expansion}},  {\em JHEP} {\bf 02} (2026)
  158, [\href{http://arxiv.org/abs/2508.13620}{{\tt arXiv:2508.13620}}].

\bibitem{Schnetz:2013hqa}
O.~Schnetz, {\it {Graphical functions and single-valued multiple
  polylogarithms}},  {\em Commun. Num. Theor. Phys.} {\bf 08} (2014) 589--675,
  [\href{http://arxiv.org/abs/1302.6445}{{\tt arXiv:1302.6445}}].

\bibitem{Golz:2015rea}
M.~Golz, E.~Panzer, and O.~Schnetz, {\it {Graphical functions in parametric
  space}},  {\em Lett. Math. Phys.} {\bf 107} (2017), no.~6 1177--1192,
  [\href{http://arxiv.org/abs/1509.07296}{{\tt arXiv:1509.07296}}].

\bibitem{Borinsky:2021gkd}
M.~Borinsky and O.~Schnetz, {\it {Graphical functions in even dimensions}},
  {\em Commun. Num. Theor. Phys.} {\bf 16} (2022), no.~3 515--614,
  [\href{http://arxiv.org/abs/2105.05015}{{\tt arXiv:2105.05015}}].

\bibitem{Schnetz:2024qqt}
O.~Schnetz and S.~Theil, {\it {Notes on graphical functions with numerator
  structure}},  {\em PoS} {\bf LL2024} (2024) 026,
  [\href{http://arxiv.org/abs/2407.17133}{{\tt arXiv:2407.17133}}].

\bibitem{HP}
O.~Schnetz, {\em {HyperlogProcedures {\sl ver.0.8}}}, 2025.
\newblock Maple package available on the homepage of the author,
  https://www.math.fau.de/person/oliver-schnetz.

\bibitem{Huang:2025ree}
R.~Huang, Q.~Jin, and Y.~Li, {\it {On the seven-loop renormalization of
  Gross-Neveu model}},  {\em JHEP} {\bf 06} (2025) 134,
  [\href{http://arxiv.org/abs/2504.00713}{{\tt arXiv:2504.00713}}].

\bibitem{Tkachov:1981wb}
F.~V. Tkachov, {\it {A theorem on analytical calculability of 4-loop
  renormalization group functions}},  {\em Phys. Lett. B} {\bf 100} (1981)
  65--68.

\bibitem{Chetyrkin:1981qh}
K.~G. Chetyrkin and F.~V. Tkachov, {\it {Integration by parts: The algorithm to
  calculate $\beta$-functions in 4 loops}},  {\em Nucl. Phys. B} {\bf 192}
  (1981) 159--204.

\bibitem{Laporta:2000dsw}
S.~Laporta, {\it {High-precision calculation of multiloop Feynman integrals by
  difference equations}},  {\em Int. J. Mod. Phys. A} {\bf 15} (2000)
  5087--5159, [\href{http://arxiv.org/abs/hep-ph/0102033}{{\tt
  hep-ph/0102033}}].

\bibitem{Smirnov:2008iw}
A.~V. Smirnov, {\it {Algorithm FIRE -- Feynman Integral REduction}},  {\em
  JHEP} {\bf 10} (2008) 107, [\href{http://arxiv.org/abs/0807.3243}{{\tt
  arXiv:0807.3243}}].

\bibitem{Maierhofer:2017gsa}
P.~Maierh\"ofer, J.~Usovitsch, and P.~Uwer, {\it {Kira\textemdash{}A Feynman
  integral reduction program}},  {\em Comput. Phys. Commun.} {\bf 230} (2018)
  99--112, [\href{http://arxiv.org/abs/1705.05610}{{\tt arXiv:1705.05610}}].

\bibitem{Wu:2023upw}
Z.~Wu, J.~Boehm, R.~Ma, H.~Xu, and Y.~Zhang, {\it {NeatIBP 1.0, a package
  generating small-size integration-by-parts relations for Feynman integrals}},
   {\em Comput. Phys. Commun.} {\bf 295} (2024) 108999,
  [\href{http://arxiv.org/abs/2305.08783}{{\tt arXiv:2305.08783}}].

\bibitem{Guan:2024byi}
X.~Guan, X.~Liu, Y.-Q. Ma, and W.-H. Wu, {\it {Blade: A package for
  block-triangular form improved Feynman integrals decomposition}},  {\em
  Comput. Phys. Commun.} {\bf 310} (2025) 109538,
  [\href{http://arxiv.org/abs/2405.14621}{{\tt arXiv:2405.14621}}].

\bibitem{Vladimirov:1979zm}
A.~A. Vladimirov, {\it {Method for Computing Renormalization Group Functions in
  Dimensional Renormalization Scheme}},  {\em Theor. Math. Phys.} {\bf 43}
  (1980) 417.

\bibitem{Chetyrkin:1980pr}
K.~G. Chetyrkin, A.~L. Kataev, and F.~V. Tkachov, {\it {New Approach to
  Evaluation of Multiloop Feynman Integrals: The Gegenbauer Polynomial x Space
  Technique}},  {\em Nucl. Phys. B} {\bf 174} (1980) 345--377.

\bibitem{Caswell:1981ek}
W.~E. Caswell and A.~D. Kennedy, {\it {A Simple Approach to Renormalization
  Theory}},  {\em Phys. Rev. D} {\bf 25} (1982) 392.

\bibitem{Chetyrkin:1982nn}
K.~G. Chetyrkin and F.~V. Tkachov, {\it {Infrared R Operation and Ultraviolet
  Counterterms in the MS Scheme}},  {\em Phys. Lett. B} {\bf 114} (1982)
  340--344.

\bibitem{CHETYRKIN1984419}
K.~Chetyrkin and V.~Smirnov, {\it {R*-Operation corrected}},  {\em Physics
  Letters B} {\bf 144} (1984), no.~5 419--424.

\bibitem{Larin:2002sc}
S.~Larin and P.~van Nieuwenhuizen, {\it {The Infrared R* operation}},
  \href{http://arxiv.org/abs/hep-th/0212315}{{\tt hep-th/0212315}}.

\bibitem{Misiak:1994zw}
M.~Misiak and M.~Munz, {\it {Two loop mixing of dimension five flavor changing
  operators}},  {\em Phys. Lett. B} {\bf 344} (1995) 308--318,
  [\href{http://arxiv.org/abs/hep-ph/9409454}{{\tt hep-ph/9409454}}].

\bibitem{Chetyrkin:1997fm}
K.~G. Chetyrkin, M.~Misiak, and M.~Munz, {\it {Beta functions and anomalous
  dimensions up to three loops}},  {\em Nucl. Phys. B} {\bf 518} (1998)
  473--494, [\href{http://arxiv.org/abs/hep-ph/9711266}{{\tt hep-ph/9711266}}].

\bibitem{Cao:2021cdt}
W.~Cao, F.~Herzog, T.~Melia, and J.~R. Nepveu, {\it {Renormalization and
  non-renormalization of scalar EFTs at higher orders}},  {\em JHEP} {\bf 09}
  (2021) 014, [\href{http://arxiv.org/abs/2105.12742}{{\tt arXiv:2105.12742}}].

\bibitem{Henriksson:2025hwi}
J.~Henriksson, F.~Herzog, S.~R. Kousvos, and J.~Roosmale~Nepveu, {\it
  {Multiloop spectra in general scalar EFTs and CFTs}},  {\em Phys. Lett. B}
  {\bf 874} (2026) 140235, [\href{http://arxiv.org/abs/2507.12518}{{\tt
  arXiv:2507.12518}}].

\bibitem{Henriksson:2025vyi}
J.~Henriksson, S.~R. Kousvos, and J.~Roosmale~Nepveu, {\it {EFT meets CFT:
  multiloop renormalization of higher-dimensional operators in general
  {\ensuremath{\phi}}$^{4}$ theories}},  {\em JHEP} {\bf 05} (2026) 005,
  [\href{http://arxiv.org/abs/2511.16740}{{\tt arXiv:2511.16740}}].

\bibitem{Collins_1984}
J.~C. Collins, {\em {Renormalization: An Introduction to Renormalization, the
  Renormalization Group and the Operator-Product Expansion}}.
\newblock Cambridge Monographs on Mathematical Physics. Cambridge University
  Press, 1984.

\bibitem{Eden:2012fe}
B.~Eden, P.~Heslop, G.~P. Korchemsky, V.~A. Smirnov, and E.~Sokatchev, {\it
  {Five-loop Konishi in N=4 SYM}},  {\em Nucl. Phys. B} {\bf 862} (2012)
  123--166, [\href{http://arxiv.org/abs/1202.5733}{{\tt arXiv:1202.5733}}].

\bibitem{Prochazka:2019fah}
V.~Proch{\'a}zka and A.~S{\"o}derberg, {\it {Composite operators near the
  boundary}},  {\em JHEP} {\bf 03} (2020) 114,
  [\href{http://arxiv.org/abs/1912.07505}{{\tt arXiv:1912.07505}}].

\bibitem{tHooft:1973mfk}
G.~'t~Hooft, {\it {Dimensional regularization and the renormalization group}},
  {\em Nucl. Phys. B} {\bf 61} (1973) 455--468.

\bibitem{Marino:2024uco}
M.~Marino and R.~Miravitllas, {\it {Trans-series from condensates}},
  \href{http://arxiv.org/abs/2402.19356}{{\tt arXiv:2402.19356}}.

\bibitem{Liu:2025bqq}
Y.~Liu and M.~Mari{\~n}o, {\it {Trans-series from condensates in the non-linear
  sigma model}},  \href{http://arxiv.org/abs/2507.02605}{{\tt
  arXiv:2507.02605}}.

\bibitem{Smirnov:2002pj}
V.~A. Smirnov, {\it {Applied asymptotic expansions in momenta and masses}},
  {\em Springer Tracts Mod. Phys.} {\bf 177} (2002) 1--262.

\bibitem{Huang:2016bmv}
R.~Huang, Q.~Jin, and B.~Feng, {\it {Form Factor and Boundary Contribution of
  Amplitude}},  {\em JHEP} {\bf 06} (2016) 072,
  [\href{http://arxiv.org/abs/1601.06612}{{\tt arXiv:1601.06612}}].

\bibitem{Huang:2022qnx}
R.~Huang, Q.~Jin, and Y.~Li, {\it {Wilson lines and boundary operators of BCFW
  shifts}},  {\em JHEP} {\bf 12} (2022) 023,
  [\href{http://arxiv.org/abs/2210.07025}{{\tt arXiv:2210.07025}}].

\bibitem{Srednicki_2007}
M.~Srednicki, {\em {Quantum Field Theory}}.
\newblock Cambridge University Press, 2007.

\bibitem{Smirnov:2019qkx}
A.~V. Smirnov and F.~S. Chukharev, {\it {FIRE6: Feynman Integral REduction with
  modular arithmetic}},  {\em Comput. Phys. Commun.} {\bf 247} (2020) 106877,
  [\href{http://arxiv.org/abs/1901.07808}{{\tt arXiv:1901.07808}}].

\bibitem{Baikov:2010hf}
P.~A. Baikov and K.~G. Chetyrkin, {\it {Four Loop Massless Propagators: An
  Algebraic Evaluation of All Master Integrals}},  {\em Nucl. Phys. B} {\bf
  837} (2010) 186--220, [\href{http://arxiv.org/abs/1004.1153}{{\tt
  arXiv:1004.1153}}].

\bibitem{Lee:2011jt}
R.~N. Lee, A.~V. Smirnov, and V.~A. Smirnov, {\it {Master Integrals for
  Four-Loop Massless Propagators up to Transcendentality Weight Twelve}},  {\em
  Nucl. Phys. B} {\bf 856} (2012) 95--110,
  [\href{http://arxiv.org/abs/1108.0732}{{\tt arXiv:1108.0732}}].

\bibitem{Kleinert:2003}
H.~Kleinert and A.~M.~J. Schakel, {\it {Gauge-Invariant Critical Exponents for
  the Ginzburg-Landau Model}},  {\em Phys. Rev. Lett.} {\bf 90} (Mar, 2003)
  097001.

\bibitem{Kleinert:2005sa}
H.~Kleinert and A.~M.~J. Schakel, {\it {Anomalous dimension of Dirac's
  gauge-invariant nonlocal order parameter in Ginzburg-Landau field theory}},
  {\em Phys. Lett. B} {\bf 611} (2005) 182--188,
  [\href{http://arxiv.org/abs/cond-mat/0501036}{{\tt cond-mat/0501036}}].

\bibitem{Irges:2017ztc}
N.~Irges and F.~Koutroulis, {\it {Renormalization of the Abelian-Higgs model in
  the $R_\xi$ and Unitary gauges and the physicality of its scalar potential}},
   {\em Nucl. Phys. B} {\bf 924} (2017) 178--278,
  [\href{http://arxiv.org/abs/1703.10369}{{\tt arXiv:1703.10369}}]. [Erratum:
  Nucl.Phys.B 938, 957--960 (2019)].

\bibitem{Georgoudis:2018olj}
A.~Georgoudis, V.~Goncalves, E.~Panzer, and R.~Pereira, {\it {Five-loop
  massless propagator integrals}},  \href{http://arxiv.org/abs/1802.00803}{{\tt
  arXiv:1802.00803}}.

\bibitem{Georgoudis:2021onj}
A.~Georgoudis, V.~Gon\c{c}alves, E.~Panzer, R.~Pereira, A.~V. Smirnov, and
  V.~A. Smirnov, {\it {Glue-and-cut at five loops}},  {\em JHEP} {\bf 09}
  (2021) 098, [\href{http://arxiv.org/abs/2104.08272}{{\tt arXiv:2104.08272}}].

\bibitem{Gracey:2016mio}
J.~A. Gracey, T.~Luthe, and Y.~Schroder, {\it {Four loop renormalization of the
  Gross-Neveu model}},  {\em Phys. Rev. D} {\bf 94} (2016), no.~12 125028,
  [\href{http://arxiv.org/abs/1609.05071}{{\tt arXiv:1609.05071}}].

\bibitem{Gracey:2025aoj}
J.~A. Gracey, A.~Maier, P.~Marquard, and Y.~Schr{\"o}der, {\it {Anomalous
  dimensions and critical exponents for the Gross-Neveu-Yukawa model at five
  loops}},  {\em Phys. Rev. D} {\bf 112} (2025), no.~8 085029,
  [\href{http://arxiv.org/abs/2507.22594}{{\tt arXiv:2507.22594}}].

\bibitem{Baikov:2016tgj}
P.~A. Baikov, K.~G. Chetyrkin, and J.~H. K\"uhn, {\it {Five-Loop Running of the
  QCD coupling constant}},  {\em Phys. Rev. Lett.} {\bf 118} (2017), no.~8
  082002, [\href{http://arxiv.org/abs/1606.08659}{{\tt arXiv:1606.08659}}].

\bibitem{Bednyakov:2023iuj}
A.~V. Bednyakov, {\it {Three-loop anomalous dimensions of fixed-charge
  operators in the SM}},  {\em Phys. Lett. B} {\bf 852} (2024) 138615,
  [\href{http://arxiv.org/abs/2312.15804}{{\tt arXiv:2312.15804}}].

\bibitem{Derkachov:1997ch}
S.~E. Derkachov and A.~N. Manashov, {\it {The Simple scheme for the calculation
  of the anomalous dimensions of composite operators in the 1/N expansion}},
  {\em Nucl. Phys. B} {\bf 522} (1998) 301--320,
  [\href{http://arxiv.org/abs/hep-th/9710015}{{\tt hep-th/9710015}}].

\bibitem{Irges:2019bzb}
N.~Irges and F.~Koutroulis, {\it {On RG flows in generalized effective field
  theory}},  {\em Nucl. Phys. B} {\bf 950} (2020) 114833,
  [\href{http://arxiv.org/abs/1907.07726}{{\tt arXiv:1907.07726}}].

\end{thebibliography}\endgroup

\end{document}